\documentclass[aps,prb,twocolumn,citeautoscript,showkeys]{revtex4-1}         
\usepackage{amsmath,amssymb,mathrsfs,bm,feynmf,setspace}
\usepackage{graphicx}     
\usepackage[tight]{subfigure}     
\usepackage{color} 
\usepackage{braket}  
\usepackage[colorlinks=true]{hyperref}  
\hypersetup{
    bookmarks=true,         % show bookmarks bar?
    unicode=false,          % non-Latin characters 
    pdftoolbar=true,        % show Acrobat
    pdfmenubar=true,        % show Acrobat 
    pdffitwindow=false,     % window fit to page when opened  
    pdfstartview={FitH},    % fits the width of the page to the window
    pdftitle={My title},    % title
    pdfauthor={Author},     % author
    pdfsubject={Subject},   % subject of the document  
    pdfcreator={Creator},   % creator of the document
    pdfproducer={Producer}, % producer of the document
    pdfkeywords={keyword1} {key2} {key3}, % list of keywords
    pdfnewwindow=true,      % links in new window
    colorlinks=true,       % false: boxed links; true: colored links 
    linkcolor=magenta, %red,          % color of internal links (change box color with linkbordercolor)
    citecolor=blue,        % color of links to bibliography
    filecolor=magenta,      % color of file links
    urlcolor=cyan           % color of external links
} 
\definecolor{darkblue}{rgb}{0.2, 0, 0.8}
\definecolor{darkgreen}{rgb}{0.2, 0.71, 0}

\newcommand{\bl}[1]{\hat #1}  

\numberwithin{equation}{section}

\newcommand{\req}[1]{(\ref{#1})} %{Eq.\thinspace(\ref{#1})}  
\newcommand{\labell}[1]{\label{#1}}
\newcommand{\etal}{{\it et.\ al.\/}}
\newcommand{\bea}{\begin{eqnarray}}
\newcommand{\eea}{\end{eqnarray}}
\newcommand{\ba}{\begin{eqnarray}}
\newcommand{\ea}{\end{eqnarray}}
\newcommand{\nn}{\nonumber \\}

\newcommand{\beq}{\begin{equation}}
\newcommand{\eeq}{\end{equation} }
\newcommand{\beqa}{\begin{eqnarray}}
\newcommand{\eeqa}{\end{eqnarray}}
\newcommand{\beqar}{\begin{eqnarray*}}
\newcommand{\eeqar}{\end{eqnarray*}}

\newcommand{\reef}[1]{(\ref{#1})}
\newcommand{\ssc}{\scriptscriptstyle}

\newcommand{\ie}{{\it i.e.,}\ }

\newcommand{\ct}{C_{T}} %{C_\mt{T}}

\newcommand{\ren}{R\'enyi\ }
\newcommand{\afour}{a^{(4d)}}
\newcommand{\sfour}{\sigma_{n}}

\newcommand{\ctt}{C_{\ssc T}}

\newcommand{\rfig}[1]{Fig.\thinspace\ref{#1}}
\newcommand{\sq}{\sigma'}
\newcommand{\amin}{\frak a_{\rm min}}
\newcommand{\aminn}{\frak a_{n}}
\DeclareMathOperator{\tr}{Tr}  

\renewcommand{\href}[2]{#2}

\begin{document}

\title{Bounds on corner entanglement in quantum critical states}  
\author{Pablo Bueno$^{\flat}$ and William Witczak-Krempa$^{\sharp}$}\vspace{0.1cm}
\affiliation{
$^\flat$ Instituut voor Theoretische Fysica, KU Leuven, Celestijnenlaan 200D, B-3001 Leuven, Belgium\\
$^\sharp$ Department of Physics, Harvard University, Cambridge MA 02138, USA 
}  
\date{\today}
\begin{abstract}     
The entanglement entropy in many gapless quantum systems receives a contribution from  
corners in the entangling surface in 2+1d. 
It is characterized by a universal function $a(\theta)$ depending   
on the opening angle $\theta$, and contains pertinent low energy information.    
For conformal field theories (CFTs), the leading expansion coefficient in the smooth limit $\theta\to \pi$ yields the stress
tensor 2-point function coefficient $C_T$.    
Little is known about $a(\theta)$ beyond that limit. Here, we show that the next 
term in the smooth limit expansion contains information beyond the 2- and 3-point correlators of the 
stress tensor. We conjecture that it encodes 4-point data, making it much richer. 
Further, we establish strong constraints on this and higher order smooth-limit coefficients.
We also show that $a(\theta)$ is lower-bounded by a non-trivial function multiplied by the central 
charge $C_T$, e.g.\ $a(\pi/2)\geq (\pi^2\log 2)C_T/6$. This bound for 90-degree corners is nearly   
saturated by all known results, including recent numerics for the interacting Wilson-Fisher quantum critical points (QCPs).
A bound is also given for the R\'enyi entropies. We illustrate our findings using $O(N)$ QCPs, free boson and Dirac fermion CFTs, 
strongly coupled holographic ones, and other models. Exact results are also given for Lifshitz quantum critical points, and for conical singularities in 3+1d.   
\end{abstract}   
\maketitle 
\singlespacing 
\tableofcontents

%%%%%%%%%%%%%%%%%%%%%%%%%%%%%% 
\section{Introduction}  
\label{sec:intro}        
In recent years, the structure of quantum entanglement in many-body systems has acquired an increasingly prominent role in 
diverse areas of physics, such as condensed matter \cite{Kitaev:2005dm,levin-wen,2005PhLA..337...22H,2008PhRvL.101a0504L,2009PhRvL.103z1601F,metlitski,swap,laflorencie},  
quantum field theory,\cite{holzhey,Calabrese1,Calabrese2,Casini1,Casini2,Casini3}  
or quantum gravity \cite{VanRaamsdonk:2009ar,VanRaamsdonk:2010pw,Bianchi:2012ev,
Balasubramanian:2013lsa,Myers:2014jia,Czech:2014wka,Headrick:2014eia,RyuTaka1,RyuTaka3,RyuTaka4,RyuTaka2}.   
The interest in the subject is perhaps not surprising given that entanglement is a fundamental property of the  
quantum realm. This being said, its recent rise in prominence can be partially attributed to the rapid development of
tools to study it in complex systems.    
In the context of condensed matter physics,   
entanglement has proven to be a powerful probe of unconventional quantum states of matter. For instance, the  
so-called topological entanglement entropy\cite{levin-wen,Kitaev:2005dm} in two spatial dimensions can be used to determine 
whether a phase is topologically non-trivial\cite{wen89}, \ie whether it possesses anyonic excitations, 
whereas conventional local order-parameters cannot.   
This diagnostic was employed to analyze realistic models for quantum spin liquids\cite{isakov,jiang12,depenbrock}.   
A closely related quantity has recently been found to organize the renormalization group flow between different 
phases and quantum critical points described by Lorentz invariant theories\cite{sinha10,CH12}. 
These results have been applied to constrain the phase diagram of models with topological phases and deconfined 
quantum critical points\cite{Swingle12,Grover12}.  

The entanglement entropy (EE) $S$ and \ren entropies \cite{Renyi1,Renyi2} $S_n$ are particularly useful measures of entanglement.  
Focusing on spatial bi-partitions, heuristically, these entropies quantify the amount of entanglement between the inside  
and outside of a given region. 
More precisely, for a spatial region $V$ and a quantum state $|\psi\rangle$, they are defined as  
\begin{align} \label{renyi}
S_n(V)&=\frac{1}{1-n}\log\, \tr \rho_V^n \,,\\ 
S(V)&=\lim_{n\to1}S_n(V)=-\tr \left( \rho_V \log \rho_V \right)\,,
\end{align}   
where $\rho_V=\text{Tr}_{\bar V}|\psi\rangle \langle\psi|$ is the reduced density matrix obtained by integrating out the degrees of freedom in the complementary (outside)
region, $\overline{V}$. 

\begin{figure}
\center
 \subfigure[]{\label{fig:triangle}\includegraphics[scale=.7]{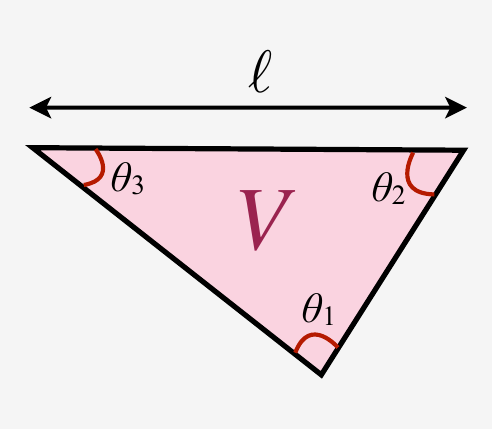}} \hspace{0.4cm}  
 \subfigure[]{\label{fig:corner}  \includegraphics[scale=.48]{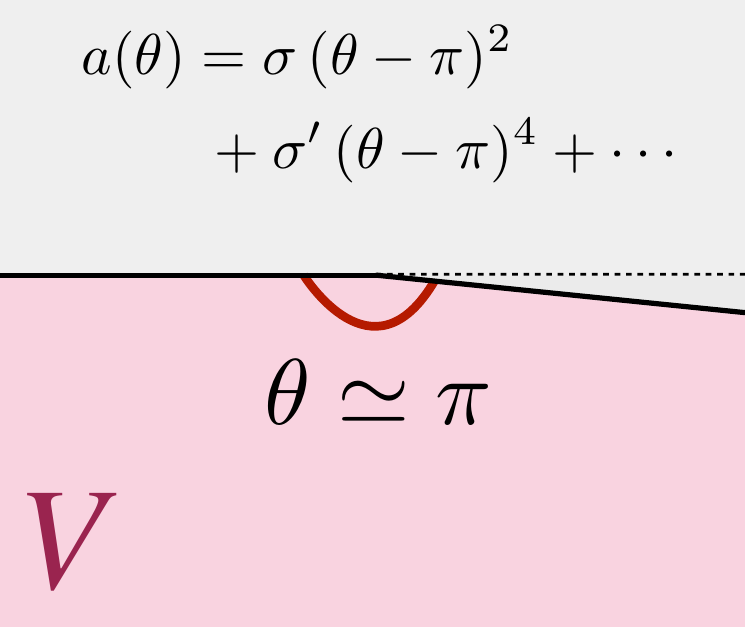}} %.49      
\caption{{\bf Entanglement from corners.} a) Corners in the entangling surface $\partial V$ with opening angles 
$\theta_i$. b) A corner with $\theta\!\simeq\!\pi$. The Taylor expansion
of the corner function $a(\theta)$ near this smooth limit yields coefficients $\sigma,\sigma',\dots$ 
that contain non-trivial information about the CFT. We have omitted the \ren index $n$ dependence. }    
\centering
\end{figure}

In this work, we shall be concerned with the entanglement and \ren entropies of strongly interacting quantum
systems described by scale invariant quantum field theories at low energy. 
Most, but not all, of our attention will be devoted to conformal field theories (CFTs).  
These possess scale and Lorentz invariance,
and for example describe the quantum critical points in the Ising and XY universality classes in $d=2+1$ spacetime
dimensions. Generally, CFTs are strongly interacting and lack quasiparticle excitations, 
unless one is dealing special theories such as free bosons or free Dirac fermions.  
We will now see how basic entanglement measures can be used to gain insights into these complex systems. 
We consider the \emph{groundstate} of a CFT in three spacetime dimensions. The corresponding \ren entropy of a region $V$ takes the generic form
\begin{align} \labell{snb} 
  S_n=B_n\frac{\ell}{\delta}-\! \sum_{i\; {\rm corners}} \! a_n(\theta_i)\log(\ell/\delta) + {\rm const.,} 
\end{align} 
where $\ell$ is a length scale characterizing the linear size of $V$, and $\delta$ is a UV cut-off: in a lattice model, this would be
the lattice spacing. $B_n$ is a regulator dependent (positive) coefficient, while the third term is finite as $\delta\to0$
and generally depends on the shape of $V$. The expansion \req{snb} holds in the limit $\ell\gg \delta$. 
While the leading divergence, corresponding to the usual 
``area'' or boundary law, is present for general entangling regions, the subleading logarithmic term appears 
only when the boundary of $V$ contains sharp corners\cite{Casini3,Casini2,Casini1,Hirata,Myers:2012vs}, each with opening angle $\theta_i$,  
as shown in \rfig{fig:triangle}.           
This contribution is characterized by a function of the corner opening angle  $a_n(\theta)$ which, as opposed to $B_n$, 
is universal (regulator-independent) and therefore encodes well-defined information about the low energy CFT. Importantly, corner contributions  
naturally arise in numerical calculations on lattice models, and since $a_n(\theta)$ is not polluted by 
UV/microscopic details it can be determined by such methods\cite{vidal09,PhysRevLett.110.135702,pitch,helmes14,devakul14,Kallin:2014oka, 
laflorencie,sahoo15,laflorencie}, and compared with quantum field theory analysis.           
      
It is instructive to begin by examining nearly smooth corners. 
On general grounds, the corner function $a_n(\theta)$ is expected to have the following expansion\cite{Casini3,Casini2,Casini1}  
near $\theta=\pi$ (see Fig.~\ref{fig:corner}) 
\begin{align} \label{expansion}
  a_n(\theta\simeq \pi) = \sigma_n\,\varepsilon^2 + \sq_n \,\varepsilon^4 + \sigma_n''\,\varepsilon^6 + \dotsb\,,
\end{align} 
where we have introduced the notation $\varepsilon\equiv (\pi-\theta)$, which will appear throughout.
We can write the expansion more succinctly as $\sum_{p=1}\sigma_n^{(p-1)}\varepsilon^{2p}$, where
$\sigma^{(0)}_n\!=\!\sigma_n$, $\sigma_n'\!=\!\sigma_n^{(1)}$, etc. 
Assuming $a_n(\theta)$ is smooth around $\theta=\pi$, the reflection property,
$a_n(2\pi-\theta)=a_n(\theta)$ --- true for pure states --- implies that only even powers of $\varepsilon$ appear in \req{expansion}. Although a proof of the analyticity of $a(\theta)$ near $\pi$ is lacking, it does hold for all the theories
for which the expansion is known (see below). 

% \begin{figure}
% \center
%    \includegraphics[scale=.5]{corner_smooth2.pdf}  
%    \caption{{\bf Entanglement from corners.} 
% A corner in the entangling surface $\partial V$ with opening angle $\theta\simeq\pi$. The Taylor expansion
% of the corner function $a(\theta)$ near this smooth limit yields coefficients $\sigma,\sigma',\dots$ 
% that contain non-trivial information about the CFT. We have omitted the \ren index $n$ dependence. 
% } 
% \labell{fig:corner}  
% \centering  
% \end{figure}  
\begin{figure*}
\center
   \includegraphics[scale=.68]{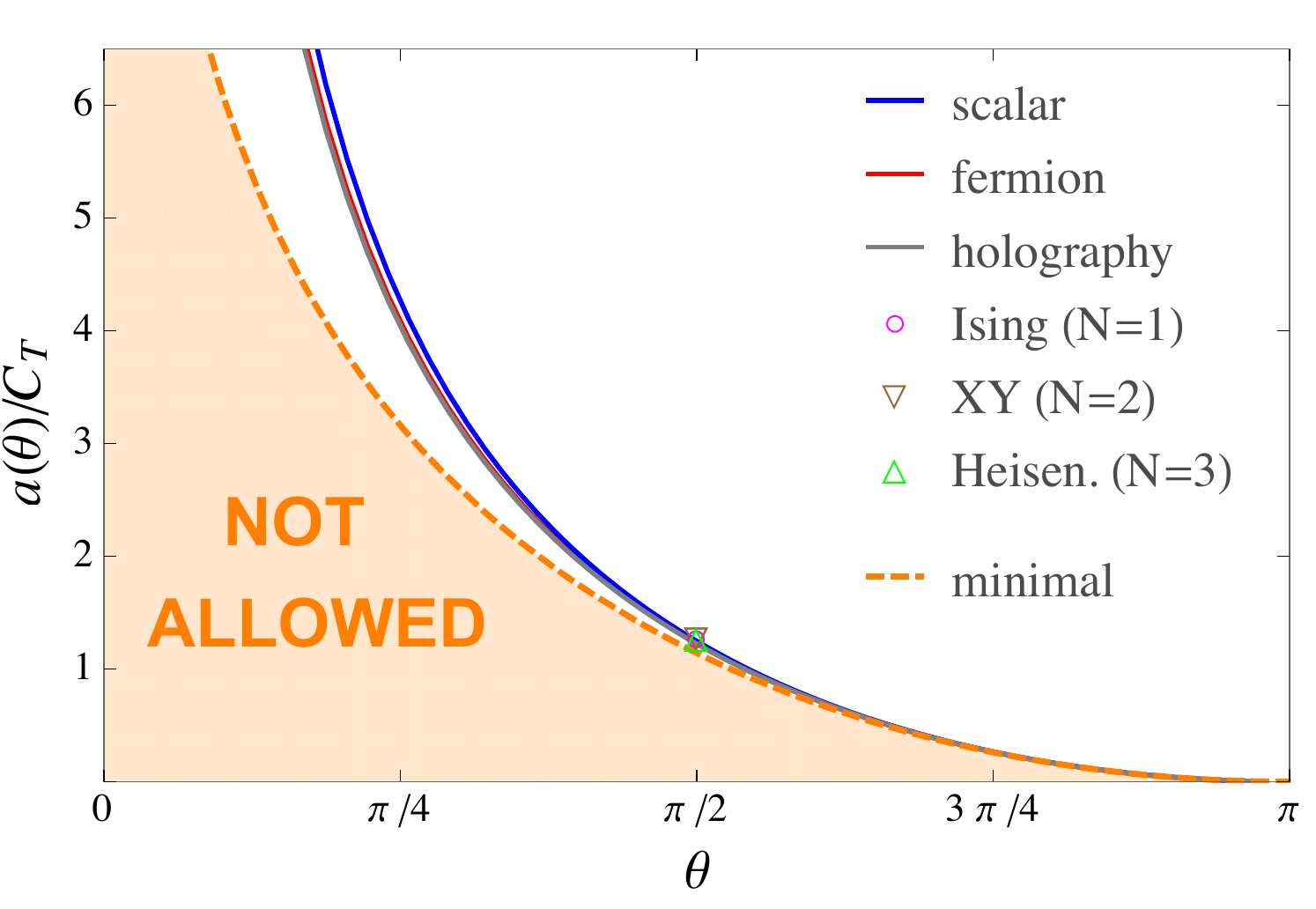}  
\caption{ {\bf Bound on the corner function.}
Corner function $a(\theta)$ normalized by the stress tensor coefficient $\ctt$ 
for a free scalar\cite{Casini1,Casini3,Osborn} (blue), a free Dirac fermion\cite{Casini1,Casini2,Osborn} (red), and a  
theory holographically dual to Einstein gravity\cite{Hirata,hong22} (gray). We also show the 
numerical results\cite{sahoo15,PhysRevLett.110.135702,Kos:2013tga,pitch,Kallin:2014oka} of the $\theta=\pi/2$ values for the $N=1,2,3$ O$(N)$ Wilson-Fisher CFTs (Table~\ref{tbl:ratio}), as well as the minimal function $\amin(\theta)/\ctt$ (orange dashed).
The shaded orange region is forbidden by the inequality $a(\theta)\geq \amin(\theta)$.   
} 
\labell{fig:bound} 
\centering     
\end{figure*}  

In recent works\cite{Bueno1,Bueno2,Bueno3} (see Refs.~\onlinecite{bueno_FdP,laflorencie} for brief reviews), the leading order coefficient in the smooth limit, $\sigma_n$, has been studied
in detail, and strong evidence \cite{Bueno1,Bueno2,Bueno3,Miao2015,Elvang,Dowker,Dowker:2015pwa,bianchi15} points to the simple relation:
\begin{align} \label{conj2}
  \sigma_n = \frac{1}{\pi}\, \frac{h_n}{n-1}\,,
\end{align}    
for any $n\!>\!0$, where $h_n$ is the conformal dimension of the so-called twist operator.\cite{holorenyi,twist,Bueno3} 
This is a line operator in $d=3$, as it has support on the entangling surface, \ie the boundary of the entangling region $V$. 
It is closely related to the swap operator used to compute the \ren entropies with quantum Monte Carlo\cite{swap}. 
In the more familiar setting of 1+1d CFTs, the twist operator is a local operator\cite{Calabrese1,Calabrese2} associated with the
endpoint of an interval. In fact, the prefactor of the logarithmically divergent term of the \ren entropy of a 
single interval\cite{holzhey,Calabrese1,Calabrese2} has the same form as the smooth-limit coefficient, Eq.~\req{conj2}.    
In the limit $n\!\to\!1$, Eq.~\req{conj2} reduces to\cite{Bueno1,Bueno2}  
\begin{align} \label{conj1}
  \sigma=\frac{\pi^2}{24} \ctt\,.   
\end{align}
%where we have omitted the subscript $1$. 
$\ctt$ is a fundamental property of the quantum critical system (CFT in this case): it is the central charge determining the 
2-point correlation function of the stress or energy-momentum tensor $T_{\mu\nu}$: 
\begin{align} \label{TT}
  \langle T_{\mu\nu}(x) T_{\eta\kappa}(0)\rangle = \frac{\ctt}{x^{2d}} \, \mathcal I_{\mu\nu,\eta\kappa}(x)\,, 
\end{align}
where $x_\mu$ is a spacetime coordinate, 
and $\mathcal I_{\mu\nu,\eta\kappa}$ is a dimensionless tensor fixed by conformal symmetry \cite{Osborn} (see Appendix \ref{ap:ttt}).
Notably, $T_{00}(x)$ corresponds to the energy density of the system; Eq.~\req{TT} determines its auto-correlation function in the groundstate. 
Eq.~\req{TT} holds in arbitrary dimensions, including $d\!=\!1+1$, where $\ctt$ is proportional to the Virasoro central charge.
Remarkably, in a recent paper by Faulkner \etal\cite{faulkner15}, Eq.~\req{conj1} was proven for general CFTs. 

% The results in that paper also imply that 
% Mezei's formula\cite{Mezei:2014zla} for the entanglement entropy of slightly deformed hyperspherical entangling surfaces in $d$-dimensional CFTs is valid for general theories and, consequently, that the generalized version of our conjecture for entangling surfaces containing hyperconical singularities\cite{Bueno4} is true for general CFTs. The generalized version of our conjecture for arbitrary values of the R\'enyi index (both for three-dimensional theories \req{conj2} and in higher dimensions\cite{Bueno4}) remains unproved in general though.
The results for the leading smooth limit coefficient, Eqs.~\req{conj2} and \req{conj1}, have also been generalized to CFTs in higher  
dimensions\cite{Bueno3,Bueno4}. In the presence of a    
(hyper)conical singularity in the entangling surface $\partial V$, the \ren entropy contains an analogous 
regulator-independent contribution characterized by a function of the cone opening angle, $a_n^{(d)}$, whose smooth limit 
expansion is analogous to \req{expansion}. The leading coefficient of such an expansion has been argued to be $\sigma_n^{\ssc(d)}=g(d)\, h_n/(n-1)$, 
where the constant $g(d)$ is known explicitly\cite{Bueno4}, and $h_n$ is the scaling dimension of a $(d\!-\!2)$-dimensional twist operator. For the special case of the EE, $n\!=\!1$, one finds $\sigma^{\ssc(d)}=\tilde g(d)\, \ctt$ where $\tilde g(d)$ is again a known $d$-dependent constant. The recent results of Ref.~\onlinecite{faulkner15}, combined with earlier analysis\cite{Bueno4,Mezei14}, complete the proof of the $n=1$ case for general CFTs in dimensions $d>3$.        

%%%%%%%%%%%%%%%%%%%%%%%%%%%%%%%%%%%%%%%%%%%%%%%%%%%%%%%%%%%%
\subsection{Main results \& outline} 
In contrast to the leading order coefficient $\sigma_n$, it is currently not known what physical information is encoded in the higher order coefficients $\{\sq_n,\sigma''_n,\dots\}$, let alone in $a_n(\theta)$ at finite angles. This work sheds light on these quantities. 
One of our main results is a lower bound on the corner entanglement function $a(\theta)$ of any CFT, 
given in Eq.~\req{amin} of \S\ref{sec:bound}, and illustrated in \rfig{fig:bound}. A lower bound for the \ren case $a_{n\neq1}(\theta)$
is also given, Eq.~\req{an_min}.  
We further prove lower bounds for the smooth limit coefficients in \S\ref{sec:smooth}; these do not follow from the bound on $a(\theta)$.
We then unravel the asymptotic properties of the high order coefficients $\sigma_n^{(p\gg1)}$ (\S\ref{sec:ssc}). \S\ref{sec:ads} studies the  
smooth limit expansion in holographic theories. In \S\ref{sec:quartic_stress}, we use the holographic results in conjunction with other information to shed light on the 
physical properties of the quartic term $\sq\,(\theta-\pi)^4$ for general theories. We show that $\sq$ contains information beyond 
the 2- and 3-point functions of the stress tensor. \S\ref{sec:lif} analyzes the structure of the corner function for a special class of Lifshitz 
quantum critical points.  
We also discuss the entanglement properties of conical singularities in higher dimensional CFTs (\S\ref{highd}).  
Finally, we give a summary and outlook in \S\ref{sec:disc}, where we provide consequential open questions. Four appendices provide details about the calculations.

\section{Lower bound on corner entanglement}   
\label{sec:bound}   
In this section we show that the strong subadditivity (SSA) property of the EE\cite{nielsen2010} imposes a non-trivial lower  
bound on the corner function, 
$a(\theta)\geq\amin(\theta)$, and we give a simple closed form expression for the minimal function $\amin(\theta)$.   
This holds for general CFTs (we are assuming \emph{gapless} theories with a finite $\ctt$, which excludes
topological quantum field theories). We also establish a bound on the corner function for general \ren index in \S\ref{sec:bound-ren}.
  
Specifically, the SSA of entanglement implies that for regions $V,V'$, the following 
inequality holds for the sum of their entanglement entropies: $S(V)+S(V')\geq S(V\cup V')+S(V\cap V')$.
Using this property, Hirata and Takayanagi showed\cite{Hirata} that $a(\theta)$ is non-negative, as well as $\partial_\theta a(\theta)\leq 0$
on $0\leq\theta\leq \pi$ ($\partial_\theta a$ is non-negative on $[\pi,2\pi]$).
Using Lorentz invariance and SSA, Casini, Huerta and Leitao\cite{Casini2} then gave the following non-trivial linear constraint:   
\begin{align} \label{ssa}  
\partial_\theta^2 a(\theta) \geq -\frac{\partial_\theta a(\theta)}{\sin \theta}\,,
\end{align}
valid for $0\leq \theta\leq 2\pi$.  
The key idea to obtain a lower bound for $a(\theta)$ is to replace the inequality in Eq.~\req{ssa} by an equality. 
This yields a linear second order differential equation, which can be readily solved 
with the appropriate boundary conditions in the smooth limit. The solution reads
\begin{align} \label{amin}
  \amin(\theta) = \frac{\pi^2\ctt}{3} \, \log\left[ 1/\sin(\theta/2) \right]\,.
\end{align}
We note that $\amin$ satisfies the reflection property,  
$\amin(\theta)=\amin(2\pi-\theta)$, expected for pure states.
It has been normalized so as to obtain the correct leading asymptotic behavior as $\theta\to\pi$:   
$\pi^2\ctt(\theta-\pi)^2/24$. 
One of our central results is that $\amin(\theta)$ provides a lower bound for the corner function of all CFTs: 
\begin{align}
  a(\theta) \geq \amin(\theta) \,, \labell{inE}
\end{align}
for all angles $0\leq\theta\leq 2\pi$.   
This inequality follows from \req{ssa}; we refer the reader to
Appendix \ref{ap:proof} for the proof, which relies on a classic result in the theory of differential inequalities. 
We point out that $\amin$ does not correspond to the corner function 
of an actual CFT since it does not have the required $1/\theta$ divergence in the sharp corner limit
$\theta\to 0\;$ Rather, it diverges only 
logarithmically, as $\log(1/\theta)$. The $\ctt$ appearing in \req{amin} should be interpreted as the stress tensor
coefficient $\ct$ corresponding to the theory   
whose $a(\theta)$ is being compared to $\amin(\theta)$. We note that a cruder version of the bound can be obtained by replacing
$\ctt$ by $24\sigma/\pi^2$ in Eq.~\req{amin}.
 
We can explicitly verify that the lower bound Eq.~\req{inE} is satisfied for CFTs both at weak and strong coupling:
Fig.~\ref{fig:bound} shows $a(\theta)/\ctt$ for
a free scalar and a free Dirac fermion\cite{Casini1,Casini2,Casini3,Bueno1}, and 
for strongly coupled holographic CFTs dual to Einstein gravity\cite{Hirata,Drukker:1999zq} --- see next section. 
We also plot the lower bound $\amin(\theta)/\ctt$. 
% This function constrains the allowed curves $a(\theta)/\ctt$ to lie 
% above it for all possible values of the opening angle. 
While the bound is  comfortably satisfied by the three theories for small opening angles, it becomes quite non-trivial 
already for $\theta\!\sim\!\pi/2$, where all theories nearly saturate it.  
This is a particularly important value of the opening angle, given that most numerical simulations that
studied $a(\theta)$ have been
performed for Hamiltonians defined on a square lattice\cite{laflorencie}. These simulations dealt with rectangular regions $V$, and 
thus obtained $a(\pi/2)$.
For 90-degree corners, our bound Eq.~\req{inE} becomes 
\begin{align} \label{bound90}    
  a(\pi/2)\geq \frac{\pi^2\log 2}{6} \ctt \approx 1.1402\, \ctt \,.
\end{align}
In both free and strongly coupled holographic theories, $a(\pi/2)/\ctt$ exceeds this bound by only a few percent,
as can be seen in \rfig{fig:bound} and Table~\ref{tbl:ratio}.  
That table also shows $a(\pi/2)/\ctt$ for the O($N$) Wilson-Fisher fixed points for $N=1,2,3$. 
These CFTs describe the quantum critical points in quantum Ising, XY, and Heinsenberg spin models, respectively. 
Remarkably, in all three cases the same ratio is found within error bars, $a(\pi/2)/\ctt=1.3(1)$, which lies not far 
above the bound Eq.~\req{bound90}. This supports the validity of the numerical data.  
We finally note that recently an analytic expression for the central charge, $\ctt=2(16\pi-9\sqrt 3)/(81\pi^3)$,
was obtained for a strongly interacting QCP with emergent supersymmetry\cite{susy}. Such a QCP is super-conformal
and can occur at the surface of a topological insulator, in which case $\ctt$ is directly proportional to
the groundstate optical conductivity\cite{susy}. Combined with Eq.~\req{amin}, one can thus
obtain a simple closed-form lower bound for $a(\theta)$ at that QCP\cite{susy}, in spite of its strongly interacting nature.     

\subsection{\ren entropy}     
\label{sec:bound-ren}
We will now generate a lower bound for $a_n(\theta)$ valid for general \ren index.\footnote{We thank Horacio Casini
for suggesting this posibility.}        
Reflection positivity of Euclidean quantum field theory leads to   
an infinite set of non-linear differential inequalities for the corner function\cite{Casini4}:
\begin{align} \label{RP}
  \det\, \big\{\partial_\theta^{j+k+2} a_n(\theta) \big\}_{j,k=0}^{M-1} \, \geq 0\, ,  
\end{align}
where $M\geq 1$ is an integer.
Although these were originally derived for integer values of $n$, we will assume they hold for all $n>0$ (by analytic
continuation). Expanding the determinants, we find that the first 2 inequalities read:
\begin{align}
  \partial_\theta ^2 a_n \geq 0\,, \label{an_convex} \\ 
  \partial_\theta^2a_n \partial_\theta^4a_n - (\partial_\theta^3a_n)^2 \geq 0\,. \label{RP2} 
\end{align}
In particular, the second equation can be recast in the form $\partial_\theta^4a_n\geq G$, with $G\geq0$, by virtue
of the convexity property, Eq.~\req{an_convex}. This form is suggestive of a lower bound, assuming we can somehow ``integrate'' the 
differential inequality. Indeed, we can apply the same methods that we used above in deriving the lower bound \req{inE} for the EE
to establish a new lower bound valid for any \ren index (see Appendix \ref{ap:bounds} for the details):   
\begin{align} \label{an_min}
  a_n(\theta) \geq \frac{h_n}{\pi(n-1)}\, (\theta-\pi)^2\,,
\end{align}
where $h_n$ is the scaling dimension of the twist operator\cite{Bueno3}, which determines the smooth limit coefficient
via Eq.~\req{conj2}. $h_n$ is the analog of $\ctt$ for the \ren entropies $n\neq 1$. 
The RHS of Eq.~\req{an_min} plays the role of $\amin(\theta)$ but for general $n$. Analogously to $\amin$, it solves the differential equation obtained by replacing   
the inequality in Eq.~\req{RP2} by an equality. Note however that for $n\!=\!1$, $\amin(\theta)$ is a stronger lower bound
since it exceeds $\sigma\,(\theta-\pi)^2$.

%We recall that $\sigma_n$ is determined by the scaling dimension of the twist operator\cite{Bueno3}, see Eq.~\req{conj2}. 
The twist dimension $h_n$ has been computed\cite{Bueno3} for all values of $n$ for the free scalar (boson) and fermion CFTs, 
as well as for strongly interacting holographic theories. 
For instance, a free complex scalar has\cite{Bueno1,Bueno3,Elvang} $h_2^{\rm cs}=1/(24\pi)$, implying a lower  
bound $a_2^{\rm cs}(\pi/2)\geq 1/96\approx 0.010$ for 90-degree 
corners. $a_2^{\rm cs}(\pi/2)$ was numerically computed\cite{Casini3} to be $0.0128$, in agreement with our bound.   
In contrast, at the O(2) interacting Wilson-Fisher quantum critical point (to which the free complex scalar flows under RG),
numerical lattice calculations have found $a_2(\pi/2)\approx 0.011$.\cite{pitch,helmes14,devakul14} The value of $h_2$
for that theory is not currently known, but given the properties of $\ctt$,\cite{Kos:2013tga} we expect it to be no smaller than $0.9h_2^{\rm cs}$.
This would yield a lower bound $a_2(\pi/2)\gtrsim 0.0094$, which is satisfied by the numerical calculations.  

It is natural to ask whether the remaining infinite set of inequalities, Eq.~\req{RP}, will yield stronger bounds,
or even an \emph{upper} bound. The answer to the former is that one can indeed get stronger bounds, 
but these require more information input compared to $h_n$, not to mention that they involve numerical solutions. 
With regards to the upper bounds, the answer is simple: those inequalities all yield lower bounds.
This essentially follows from the fact that the reflection positivity inequalities 
are encoded by the determinant of the same larger and larger matrix (see Appendix \ref{ap:bounds}).  

Eq.~\req{an_min} immediately implies that the quartic coefficient $\sigma_n'$ is positive 
for all $n$, a fact that we will now rederive, along with other bounds on the smooth-limit expansion coefficients.

%%%%%%%%%%%%%%%%%%%%%%%%%%%%%%%%%%%%%%%%%%%%%%%%%%%%%%%%%%%%%%%%%%%%%%%%%%%%%%%%%%%%%%%%%%
\begin{table*} %[b!]
  \centering
  \begin{tabular}{c||c|c|c|c|c|c|c}   
     & lower bound &  Ising ($N\!=\!1$) & XY $(N\!=\!2)$  & Heisen.\ ($N\!=\!3$) & scalar  & Dirac fermion & AdS/CFT  \\
    \hline \hline
    $a(\pi/2)/\ctt$ & 1.140 & 1.3(1)\, \cite{sahoo15,PhysRevLett.110.135702,Kos:2013tga} & 1.3(1)\, \cite{pitch,Kos:2013tga} 
& 1.3(1)\, \cite{Kallin:2014oka,Kos:2013tga} & 1.245\, \cite{Casini1,Casini3,Osborn} & 1.226\,  
\cite{Casini1,Casini2,Osborn} & 1.222\, \cite{Hirata,hong22}  \\
$a(3\pi/4)/\ctt$ & 0.260 &- & - & - & 0.265\,\cite{Casini1,Casini3,Osborn} & 0.264\, \cite{Casini1,Casini2,Osborn} & 0.264\, \cite{Hirata,hong22}  \\ 
   % $a_2(\pi/2)/\sigma_2$ & 2.47 & 
  \end{tabular}
  \caption{Ratio $a(\theta)/\ctt$, with $\theta=\pi/2,3\pi/4$, for different critical theories described by CFTs, including the Wilson-Fisher
quantum critical points. 
The first entries are the lower bounds we derived, Eq.~\req{amin}. } 
\label{tbl:ratio}    
\end{table*}    

% \subsection{Potential bound at R\'enyi index $n<1$}
% In Ref.~\onlinecite{Bueno3} it was shown that $a_n(\theta)$ diverges like $1/n^2$ for any
% $0<\theta<\pi$, and tends to a finite function $a_\infty(\theta)$ as $n\to\infty$. 
% These results suggest that at fixed $\theta$, $a_n(\theta)$ decreases monotonically with $n$.
% If this were to hold, we could use the bound Eq.~\req{inE} to establish that $a_n(\theta)>\amin(\theta)$
% for any $n<1$.  

%%%%%%%%%%%%%%%%%%%%%%%%%%%%%%
\section{Constraining the expansion about the smooth limit}
\label{sec:smooth}  
In this section we use the constraints found by Casini and Huerta on the corner functions $a_n(\theta)$ from reflection positivity\cite{Casini4} and 
SSA of EE\cite{Casini1,Casini2,Casini3} to establish bounds on the  
smooth-limit expansion coefficients $\sigma_n^{(p)}$.

\subsection{Bounds for general \ren index} 
\label{ssineq}
For general \ren index, we can use the infinite set of non-linear inequalities for the corner function $a_n(\theta)$ coming from reflection positivity\cite{Casini4},
Eq.~\req{RP}, to constrain  
the smooth-limit coefficients. Substituting the expansion Eq.~\req{expansion} into the first few inequalities, we find 
\begin{align}  \label{smooth-Ren-ineq}  
  \sigma_n' &\geq 0\, ,\\ %\nn
  \sigma_n'' &\geq \frac{2}{5}\, \frac{(\sq_n)^2}{\sigma_n} \, ,\nn
  \sigma_n''' &\geq \frac{15}{28}\, \frac{(\sigma_n'')^2}{\sq_n} \, ,\nn
   \sigma_n^{(4)} &\geq \frac{45(\sigma_n'')^3-168\sigma_n'\sigma_n''\sigma_n'''+392\sigma_n(\sigma_n''')^2}{126[5\sigma_n\sigma_n''-2(\sigma_n')^2]}\, . \nonumber
%\nn  & \;\; \vdots 
\end{align} 
Similar inequalities can be obtained for $\sigma^{(p)}$, $p>4$, but we omit them as they are not particularly illuminating.  
We have been able to show that Eq.~\req{smooth-Ren-ineq} leads to --- see Appendix \ref{ineq56} --- 
\begin{align}  
  \sigma_n,\sq_n,\sigma''_n,\sigma_n''',\sigma_n^{(4)} \geq 0\,. 
\end{align}  
%\sigma_n^{(5)}
Namely, the first five expansion coefficients must be positive for any $n>0$.  
We suspect that this holds
for \emph{all} the coefficients, however, the inequalities become sufficiently complicated
at higher order, $\sigma^{(p>4)}$, that we cannot at present prove this claim. 
This being said, in section \ref{sec:ssc}, we will argue that the coefficients $\sigma_n^{(p)}$ are strictly positive at 
sufficiently large $p$. It is not unreasonable then to expect that the positivity property extends to intermediate values
of $p$. 
% \begin{align}\label{ssc}
% \sigma_n^{(p)}\rightarrow \frac{2}{\pi^{2p+3}}\kappa_n
% \end{align}  
% as $p\rightarrow \infty$, where $\kappa_n$ is the leading coefficient in the $a_n(\theta)$ expansion around $\theta=0$ --- \ie $a_n(\theta\rightarrow 0)\sim \kappa_n/\theta$. $\kappa_n$ is constrained to be positive by the general inequality $a(\theta)\geq0$. Hence, we find that the smooth expansion coefficients $\sigma_n^{(p)}$ will also be positive for sufficiently large values of $p$,
%  \begin{align}\label{pverybig}
%  \sigma^{(p\gg 1)}_n\geq 0\, ,
%  \end{align}  
%  as long as \req{ssc} is satisfied. Besides, observe that the corner coefficients of all the known theories seem to satisfy a relation of the form $\sigma_n^{(p)}\geq \sigma_n^{(p+1)}$ for all values of $p$ which, along with \req{pverybig}, would immediately imply $\sigma_n^{(p)}\geq 0$ for all values of $p$.
 
% It is also natural to expect that the analogous coefficients in the almost-smooth cone expansions for higher-dimensional theories satisfy the inequality $\sigma_n^{(d),(p)}\geq 0$ as well. We return to this point in section \ref{highd}.

\subsection{Bounds for the entanglement entropy} %from strong subadditivity of entanglement} 
%We can show a stronger inequality for $\sq_n$ at $n=1$ 
If we substitute the smooth limit expansion of $a(\theta)$, \req{expansion}, into the differential inequality obtained for the EE Eq.~\req{ssa},  
we obtain the following constraint for the quartic coefficient:  
\begin{align} \labell{inequ}
  \sq\geq \frac{\sigma}{24}\, .
\end{align}
%In particular, this implies that $\sq$ is positive, in agreement with the analysis above. 
We can combine this inequality with
the result \req{conj1} relating $\sigma$ and $\ctt$ to obtain:
\begin{align} \label{sigp-constr}
  \frac{\sq}{\ctt} \geq \frac{\pi^2}{24^2} \approx 0.0171\, .
\end{align}
We can further combine this inequality with the ones obtained above, Eq.~\req{smooth-Ren-ineq}, to generate bounds
for the higher order coefficients:
\begin{align}
  \frac{\sigma''}{\ctt} &\geq \frac{\pi^2}{34\thinspace 560} \approx 2.86\times 10^{-4}\, , \nn
  \frac{\sigma'''}{\ctt} &\geq \frac{\pi^2}{3\thinspace 870\thinspace 720} \approx 2.55\times 10^{-6} \, .
 \label{sigpp-constr} 
\end{align}

It would be interesting to see whether stronger constraints can be derived for $\sigma'',\sigma''',\dots$, 
compared to what arises from the general-$n$ inequalities \req{smooth-Ren-ineq}. One possibility 
consists of looking at the smooth-limit coefficients of $\amin(\theta)$.
%In the next subsection we give an argument for why this might be the case.
%\comment{paragraph on $\amin$, fig. \ref{fig:bound} and implications on CM calculations, table with the values of $\amin(\pi/4,\pi/2,3\pi/4)$ compared to holo and free fields?}
%In any event, \req{inE} is already a rather non-trivial constraint...
The expansion of this function near $\pi$ is as follows:
\begin{align} \label{min-expansion}   
 \frac{\amin(\theta)}{\ctt} = \frac{\pi^2}{24} \varepsilon^2 + \frac{\pi^2}{24^2} \varepsilon^4+\frac{\pi^2}{8640} \varepsilon^6
 + \mathcal O\left(\varepsilon^{8}\right)\, .
 %+\frac{17 \pi ^2 }{1935360} (t-\pi )^8 
\end{align}
We thus see that the second coefficient exactly corresponds to the lower bound for $\sq$ obtained above. 
Could it be that the other coefficients provide lower bounds for $\sigma'', \sigma'''$, etc?
For instance, is it true that any CFT will have $\sigma''/\ctt \geq \pi^2/8640$?
Should these hold, they would be stronger lower bounds than the ones obtained above. 
In Appendix \ref{ap:exact}, we show that all the expansion coefficients of $\amin$ are positive,
and give them in closed-form.
The coefficients that have been computed for holographic CFTs, and free CFTs indeed lie above those of the minimal
function, as shown in Fig.~\ref{fig:sigmap}. In this figure we also observe that the smooth-limit coefficients
of these theories seem to behave as $\sigma^{(p\gg 1)}\rightarrow A/ \pi^{2p}$, where $A$ is a theory-dependent constant.
We analyze this behavior more closely in the following section.

\begin{figure}
\center
   \includegraphics[scale=.37]{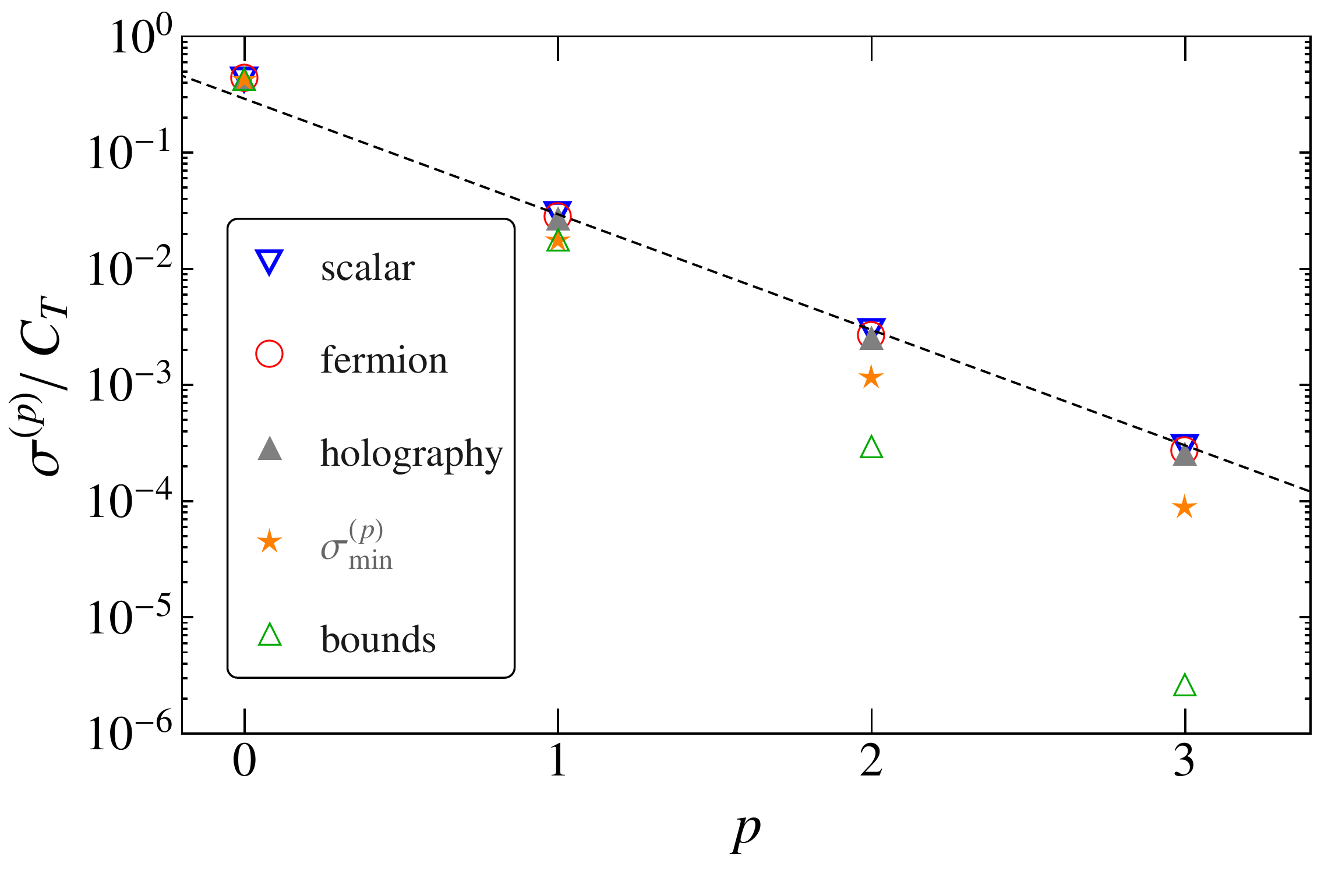}
\caption{{\bf Smooth limit coefficients and bounds.}
Log-linear plot of the first four smooth limit coefficients $\sigma^{(p)}$, $p\leq 3$, normalized by $\ctt$ for the free scalar,
free fermion, and holographic CFTs dual to Einstein gravity. $\sigma_{\rm min}^{(p)}$ are the coefficients of
the minimal function $\amin(\theta)$. We also plot the bounds obtained in \req{sigp-constr} and \req{sigpp-constr}.  
The dashed line corresponds to the exponential scaling $A/\pi^{2p}$, where the properties of $A$ are revealed in \S\ref{sec:ssc}
and \rfig{fig:smoothsharp}.  
} 
\labell{fig:sigmap}  
\centering   
\end{figure}   

\section{Smooth-sharp connection}
\label{sec:ssc}  
%At the end of section \ref{ssineq} and 
As shown in Fig.~\ref{fig:sigmap}, the coefficients of the Taylor expansion around  
$\theta=\pi$ decay as $\pi^{-2p}$ for growing values of $p$, both for free fields and holographic theories.
Here we will argue that this exponential decay occurs generally for the corner function (at any \ren index), 
and even for 
conical singularities in higher dimensions. The starting point of our discussion is, perhaps surprisingly,
as far away from the smooth limit as one can go. In the sharp corner limit, the corner entanglement function diverges
as $1/\theta$:
\begin{align} \label{sharp}
  a_n(\theta\to 0)= \frac{\kappa_n}{\theta}\,,
\end{align}
where $\kappa_n>0$ is the so-called sharp limit coefficient. This implies that the Taylor expansion about $\pi$
%diverges as $\theta$ vanishes, and 
has at most a radius of convergence of $\pi$ due to the divergences at $\theta=0,2\pi$. We claim that 
for any CFT, the corner function has this radius of expansion precisely equal to $\pi$. In other words,
the smooth limit expansion is as well-behaved as possible. In particular, this implies that the smooth limit expansion must
encode the divergence as $\theta\to 0$. It is not hard to see that the geometric series $\sum_{p=0}(\pi-\theta)^{2p}/\pi^{2p}$
yields $\pi/(2\theta)$, which has the desired divergence. Thus we deduce that the expansion
coefficients of corner function will share the same asymptotics: 
\begin{align} \label{sigp-asym}
  \sigma^{(p\gg 1)}_n \to \frac{2\kappa_n}{\pi^{2p+3}}\,,
\end{align}  
which yields the correct weight for the pole at zero opening angle, Eq.~\req{sharp}.
In Fig.~\ref{fig:smoothsharp}, we show that the asymptotic scaling given in Eq.~\req{sigp-asym} already
becomes a good approximation at $p\!=\!2$ for the free scalar\cite{Casini1}, free Dirac fermion\cite{Casini1}$^,$
\footnote{Regarding the free scalar and Dirac fermion, $\sigma^{(p)}$ are given in Ref.~\onlinecite{Casini1}   
for $0\leq p\leq 3$. We are grateful to Horacio Casini for sharing with us the unpublished results for higher $p$,
as shown in \rfig{fig:smoothsharp}.}, and holographic theories. Indeed,        
the ratio $\sigma^{(p)}\pi^{2p+3}/2\kappa$ obtained for the EE deviates from unity by less than
$2\%$ for $p\geq 2$. In addition, Fig.~\ref{fig:smoothsharp} shows the coefficients for the so-called Extensive Mutual
Information model\cite{Casini:2008wt,Swingle:2010jz,Casini:2005rm}, and the cone function $a^{4d}(\theta)$ valid for \emph{all} CFTs in 3+1 dimensions (see Section~\ref{highd}). 

We thus see that knowledge of the first few smooth limit coefficients can give a good estimate of the
sharp limit coefficient $\kappa$ via the smooth-sharp relation Eq.~\req{sigp-asym}. Conversely, knowledge of $\kappa$ can be used to estimate the smooth limit coefficients even for relatively small values of $p$.
\begin{figure}    
	\center 
	\includegraphics[scale=.48]{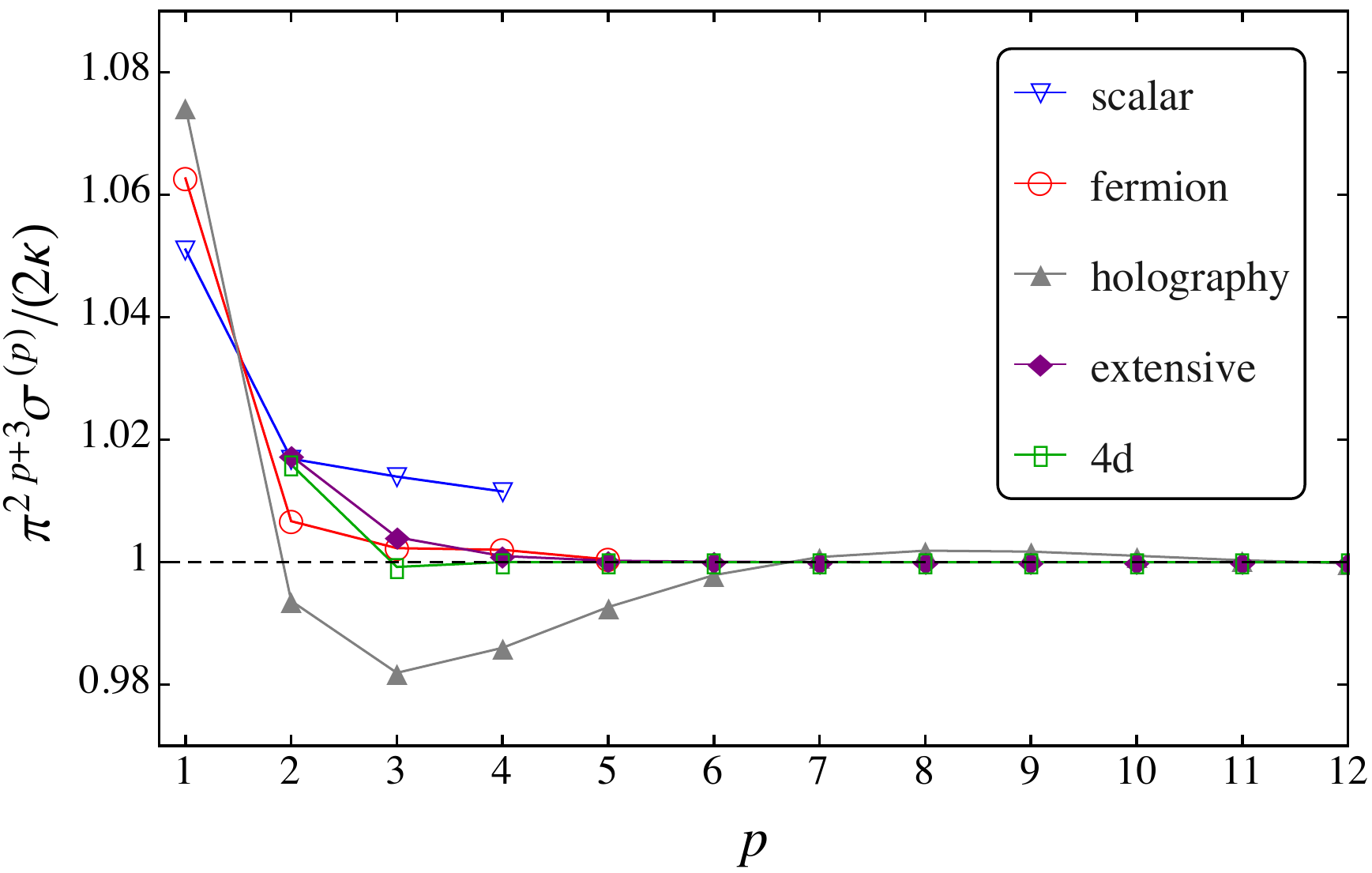} 
	\caption{ {\bf Smooth-sharp connection.}
		Smooth limit coefficients $\sigma^{(p)}$ normalized by the sharp limit coefficient $\kappa$ 
(times $2/\pi^{2p+3}$). We show results for a free scalar (blue), 
a free Dirac fermion (red), holography dual to Einstein gravity (gray), the extensive mutual information model (purple), 
and the conical $d=4$ function (green). 
All the ratios approach $1$ as $p$ grows, in agreement with Eq.~\req{sigp-asym}.   
	} 
	\labell{fig:smoothsharp} 
	\centering  
\end{figure}

\section{Smooth limit expansion in AdS/CFT} 
\label{sec:ads}
\begin{figure}
\center
 \subfigure[]{\label{fig:h0}\includegraphics[scale=.45]{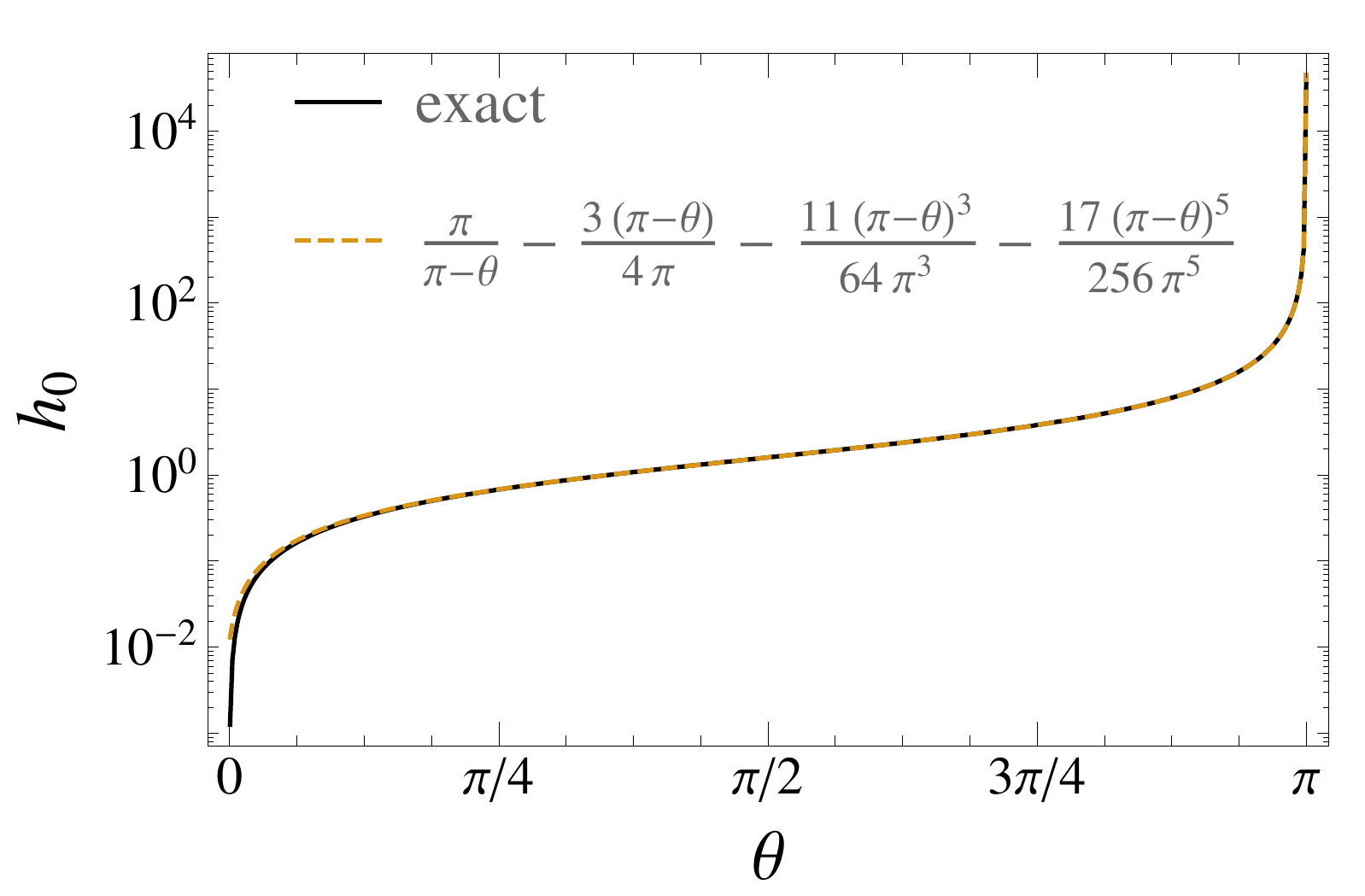}} %\hspace{1cm} 
 \subfigure[]{\label{fig:aE}  \includegraphics[scale=.45]{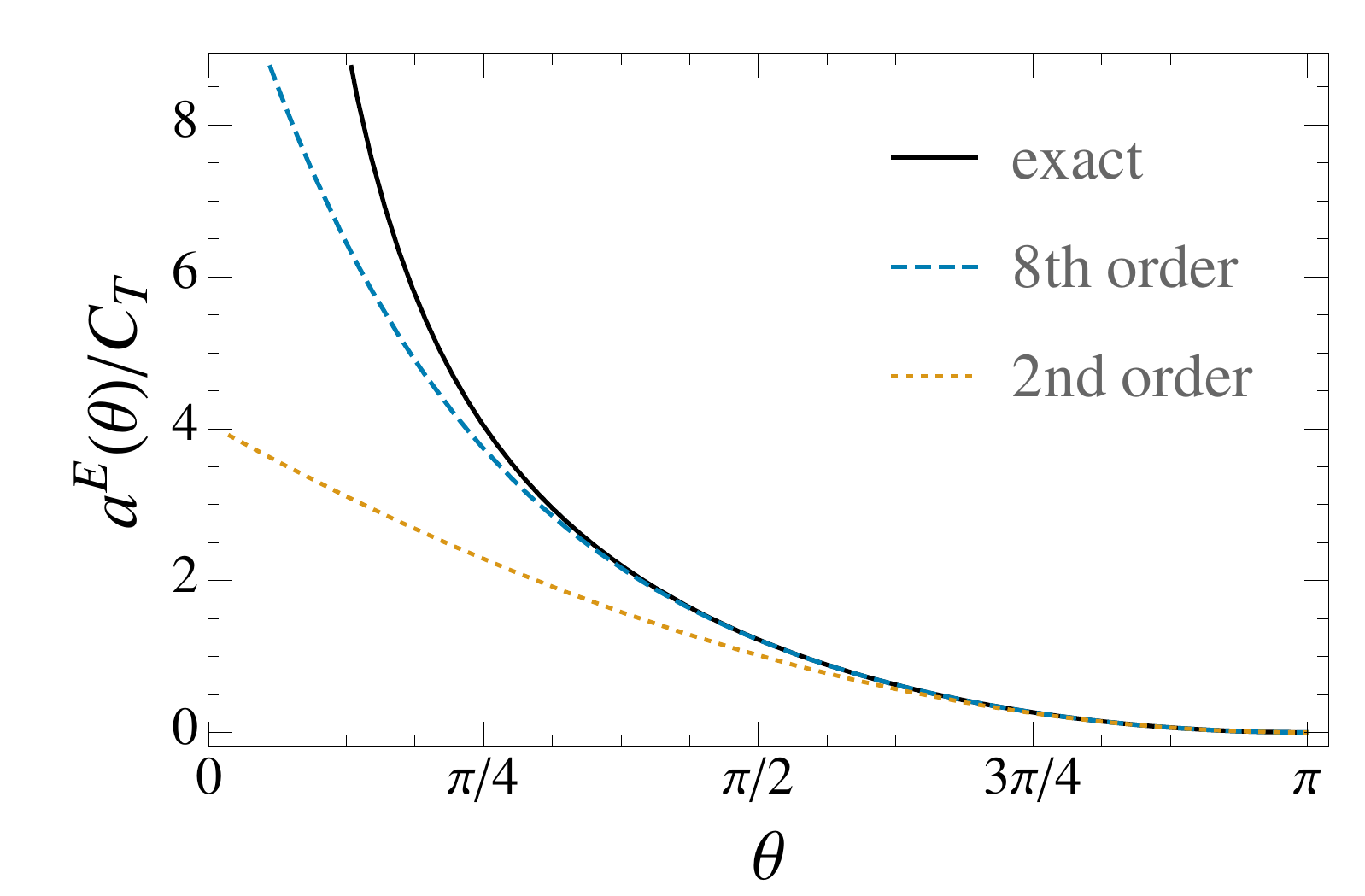}} %.49    
\caption{{\bf Holographic corner function.} a) Log-linear plot of the holographic depth parameter $h_0$ versus the 
corner opening angle $\theta$. The truncated 5th order expansion near
 $\theta=\pi$ (dashed) accurately captures the angle dependence even away from $\pi$ (deviations only
 become apparent as $\theta\!\to\!0$). b) Corner function $a^E(\theta)$, normalized by $\ctt$, 
 and the corresponding approximations to $\ell$-th order in $(\theta-\pi)^\ell$.}  
\labell{hol_aE} 
\centering
\end{figure}

In this section, we study the corner function of strongly interacting CFTs described in terms of a higher-dimensional theory of gravity via the
holographic Anti de Sitter (AdS) / CFT correspondence\cite{Maldacena}. The results we derive for the expansion in the smooth limit $\theta\!\to\!\pi$
will be used to deduce general properties of the quartic coefficient $\sq$ in Section \ref{sec:quartic_stress}.  
We focus on theories with the simplest holographic bulk description, corresponding to pure Einstein gravity
in the 3+1 dimensional AdS spacetime. The holographic minimal surface prescription of Ryu and 
Takayanagi\cite{RyuTaka1,RyuTaka2,RyuTaka3,RyuTaka4} allows one to compute the EE associated with a region $V$ in the boundary theory 
by extremizing the area of a surface extending from the entangling surface $\partial V$ to the holographic bulk. 
In particular, one can consider a wedge-shaped region with a corner of opening angle $\theta$.
The resulting corner function was obtained implicitly via \cite{Drukker:1999zq,Hirata}:
\begin{align}  \label{aE}  
  a^E(\theta)=\frac{L^2}{2G} \int_0^\infty\! ds \left[\, 1-\sqrt{\frac{1+h_0^2\,(1+s^2)}{2+h_0^2\,(1+s^2)}}\, \right]\, ,
\end{align}
where $L$ is the AdS radius, $G$ is the bulk's gravitational constant, and $h_0$ is independent of $s$.
We note that this integral can be evaluated exactly in terms of elliptic functions\cite{Fonda:2014cca,Fonda:2015nma}.
$h_0$ is implicitly determined by the corner opening angle $\theta$ via:
\begin{align} \label{h0def}
  \theta = \int_0^{h_0} \!dh\, \frac{2h^2}{\sqrt{1+h^2}} \frac{1}{\sqrt{h_0^2-h^2}} \frac{1}{\sqrt{h_0^2(1+h_0^2)^{-1}+h^2}}\, .
\end{align}
Heuristically, $h_0$ is the proportionality constant that determines how deep the minimal-area surface extends into the bulk
along its spine, when moving away from the tip of the corner. In the very sharp limit $\theta\!\to\!0$, 
the Ryu-Takayanagi minimal surface is very shallow, resulting in a small $h_0$.
$h_0$ is plotted in \rfig{fig:h0}, and we see that
it vanishes as $\theta\to 0$, and diverges as $\theta\to\pi$, as expected.   
%Now, we observe that the smooth limit $\theta\to\pi$ corresponds to $h_0\to\infty$.
In order to obtain an expansion for $a^E(\theta)$ 
near $\pi$, we first need to expand the RHS of Eq.~\req{h0def} about
$h_0=\infty$. However, a direct Taylor expansion of the integrand of \req{h0def} is not well-defined.
We circumvent this issue by expanding the last factor, $1/\sqrt{h_0^2(1+h_0^2)^{-1}+h^2}$, and keeping the problematic one, 
$1/\sqrt{h_0^2-h^2}$, intact. We can then perform the $h$ integral exactly, and finally expand in powers of
$1/h_0$. We find: 
\begin{align}
  \frac{\pi-\theta}{\pi} = \frac{1}{h_0} - \frac{3}{4h_0^3} + \frac{61}{64 h_0^5} - \frac{359}{256 h_0^7} 
 + \mathcal O\! \left(\frac{1}{h_0^9}\right),
\end{align}
which can be inverted to yield:
\begin{align}
  \!\! h_0 = \frac{\pi}{\varepsilon}- \frac{3}{4}\! \left(\frac{\varepsilon}{\pi}\right) - \frac{11}{64}\! \left(\frac{\varepsilon}{\pi}\right)^{\!3}
  - \frac{17}{256}\!\left(\frac{\varepsilon}{\pi}\right)^{\! 5}  
  +\mathcal O(\varepsilon^7),
\end{align}
where $\varepsilon=\pi-\theta$. 
We finally substitute this expansion into the integral for $a^E(\theta)$, Eq.~\req{aE}, and get:  
\begin{align}
  a^E(\theta) = \frac{L^2}{2G}\!\left[ \frac{\varepsilon^2}{4\pi} +   
    \frac{5\varepsilon^4}{32\pi^3} + \frac{37\varepsilon^6}{256\pi^5} %+ \frac{585\varepsilon^8}{4096\pi^7}
 +\dotsb \right].
\end{align}
Using $\ctt=3L^2/(\pi^3 G)$, we can express the first smooth-limit expansion coefficients as:
\begin{align}
  \frac{\sigma}{\ctt} &= \frac{\pi^2}{24}\, , \qquad \qquad \qquad  \frac{\sq}{\ctt} = \frac{5}{192}\, , \nn
  \frac{\sigma''}{\ctt} &= \frac{37}{1536\pi^2}\, , \qquad \qquad \frac{\sigma'''}{\ctt} = \frac{195}{8192\pi^4}\, , \nn
    \frac{\sigma^{(4)}}{\ctt} &= \frac{3133}{131072\pi^6}\, , \quad \qquad \frac{\sigma^{(5)}}{\ctt} = \frac{25233}{1048576\pi^8} \, .
    \label{sigp-ads}
\end{align}
Note that all these are positive, and are described by the formula $r/\pi^{2(p-1)}$, where $r$ is a rational  
number. We have verified that the first 20 coefficients satisfy this simple form; it 
would be interesting to obtain an analytical expression for all the coefficients.  
\rfig{fig:sigmap} shows that the bounds on $\sigma^{(p\leq3)}$ are comfortably satisfied.  
In addition, we have verified that these coefficients lie close to the coefficients of the closed-form approximate
form for $a^E(\theta)$ that we have previously derived\cite{Bueno3}, which is further support for the latter.

As noted in the introduction, there exists strong evidence that $\sigma$ is determined by $C_T$.
It is then natural to ask what physical properties determine the other coefficients $\sigma',\sigma''$, etc? 
We now turn to this question. 

\section{Quartic coefficient and stress tensor correlators} \label{sec:quartic_stress}   

The main goal of this section is to investigate what properties determine the quartic coefficient $\sq$.
Since the leading coefficient $\sigma$ is fixed by the 2-point function of the stress tensor,
one might expect that the next coefficient depends on the 3-point function.
The 3-point function of the stress tensor is highly constrained by conformal invariance
and only depends on two numbers:
$\ctt$, which determines the 2-point function, and an independent constant, $t_4$, see appendix \ref{ap:ttt}. 
% The constant $t_4$ appears in the groundstate 3-point function 
% $\langle T_{\alpha\beta}T_{\gamma\delta}T_{\eta\kappa}\rangle$, appendix \ref{ap:ttt}. 
If $\sq$ were to depend solely on $\ctt$ and $t_4$, it would have to be on a \emph{linear} combination of the two constants.
This is because $a(\theta)$ is additive for free CFTs, just as $\ctt$ and the product $t_4 \ctt$.  
Indeed, for $n_s$ complex free scalars and $n_{\rm f}$ 2-component Dirac fermions\cite{Osborn,buchel}: 
\begin{align} 
  \ctt &= \frac{3}{16\pi^2} (n_{\rm cs}+n_{\rm f})\,, \nn
  t_4 &= 4\,\frac{n_{\rm cs}-n_{\rm f}}{n_{\rm cs}+n_{\rm f}} \,.
\end{align}
This implies that without loss of generality we can work with an Ansatz of the form
$\sq=c_1+c_2\ctt +c_3 \ctt t_4$. Actually, $c_1=0$ due to the additive property of the free field CFTs. 
We can also fix $c_2$ by virtue of our findings in \S\ref{sec:ads}:
\begin{align} \label{ansatz} 
  \frac{\sigma'}{\ctt} = \frac{5}{192} + c_3 \, t_4 \,,
\end{align}
where we have used the value of $\sq$ obtained above for Einstein holography, Eq.~\req{sigp-ads}, for which    
$t_4=0$\cite{hofman,buchel}. For the complex scalar and Dirac fermion free CFTs, Casini and Huerta numerically found\cite{Casini1} 
\begin{align} 
  \sq_{\rm cs}/\ctt = 0.0287088\,, \qquad \sq_{\rm f}/\ctt = 0.0263940\,,  
\end{align}
where the uncertainties are on the last digit. 
We note that both these ratios exceed the holographic one, $5/192\approx 0.0260417$. 
Now, the complex scalar has a positive $t_4=4$, 
implying that $c_3\!>\!0$ for \req{ansatz} to hold. However, this is inconsistent with the fact that the fermion has
a negative $t_4=-4$ but $\sq_{\rm f}/\ctt >5/192$. 
Our analysis thus demonstrates that $\sq$ is not fully determined by the 2- and 3-point function data of the 
stress tensor. 

Interestingly, Ref.~\onlinecite{Miao2015} claimed that for a class of CFTs dual to holographic bulk theories 
with higher derivative terms, the quartic coefficient is entirely determined by $\ctt$
and $t_4$, and thus given by Eq.~\req{ansatz}. As we have shown, this stands in contrast to the general case where additional
information enters in $\sq$.
The specific value of the constant $c_3$ in Eq.~\req{ansatz} was not provided in Ref.~\onlinecite{Miao2015}, 
but it would be interesting to investigate  
how close these holographic theories approach the bounds we have derived in \S\ref{sec:smooth}. Alternatively,
our lower bounds would translate into bounds on the coupling of the higher derivative holographic theories. 

\subsection{4-point data}   
Our analysis above showed that 2- and 3-point function data is insufficient to characterize $\sq$.
We claim that in addition to this data, \ie $\ctt,t_4$, information about the stress tensor 4-point function is needed
to fix $\sigma'$. 
However, this is more difficult to show and test because the conformal symmetry is not sufficient to
fix the coordinate dependence of $\langle TTTT\rangle$, in contrast to 2- and 3-point correlators.
In other words, the 4-point correlator will contain undetermined functions of the conformal 
cross ratios (see below). These functions are theory dependent. To illustrate the point, it is sufficient to consider 
the 4-point function of a scalar operator $O(x)$ with scaling dimension $\Delta$:
\begin{align}
  \langle O(x_1)O(x_2) O(x_3)O(x_4) \rangle = \frac{F(u,v)}{x_{12}^{2\Delta} x_{34}^{2\Delta}}\,,
\end{align}
where $x_{ij}=x_i-x_j$, and the conformal cross ratios are $u=x_{12}^2x_{34}^2/(x_{13}^2x_{24}^2)$
and $v=x_{14}^2x_{23}^2/(x_{13}^2x_{24}^2)$. $F(u,v)$ is an arbitrary theory-dependent function. 

We leave for future work the investigation of the specific role that the $T_{\mu\nu}$ 4-point correlator plays
in determining $\sigma'$. It is natural to expect that the higher point functions of the stress tensor 
describe the higher order coefficients. 

\section{Lifshitz quantum critical points}
\label{sec:lif}

Up to this point our analysis was focused on CFTs, but it is natural to ask whether our results hold for a larger class
of quantum critical theories. To illustrate that this is indeed the case, we will analyze the properties of 
a special class of Lifshitz quantum critical points in $d\!=\!2+1$. These have a dynamical exponent $z=2$,  
and have scale-invariant groundstate wavefunctions\cite{ardonne}. In fact, their equal-time correlation functions are determined by  
a parent $2$-dimensional CFT. 
For this reason they are often called ``conformal'' quantum critical points. The quotes appear since these theories do
not possess the full $d=3$ spacetime conformal symmetry, but rather a time-independent version.   
For example, this class of Lifshitz theories describes lattice models such as the
quantum dimer\cite{rk} and the quantum eight-vertex\cite{ardonne} models. In the former case, on a square lattice,
the following quantum critical theory emerges at low energy: 
\begin{align} \label{L-lif}
  \mathcal L = \frac{1}{2} (\partial_\tau \varphi)^2 + \frac{\kappa^2}{2}(\partial_x^2\varphi +\partial_y^2\varphi)^2\,,
\end{align}
where $\varphi$ is a real scalar field.
Eq.~\req{L-lif} has a $d=2$ free boson CFT describing its equal-time correlators (it has Virasoro central charge $c=1$).    

When the entangling surface contains corners, the EE has a logarithmic subleading term\cite{fradkin} just as for CFTs, Eq.~\req{snb}.
Owing to the spatial conformal invariance of the wavefunction, the corresponding corner function of such theories can be 
given in closed-form\cite{fradkin}:  
\begin{align}\label{lif}
  a(\theta) = \frac{c}{12} \, \frac{(\theta-\pi)^2}{\theta(2\pi-\theta)}\,,
\end{align}
where $c$ is the central charge of the parent $d=2$ CFT. As pointed out in Ref.~\onlinecite{Bueno3}, 
this is the simplest $\theta$-dependence compatible with the expected behavior in the limits $\theta\to 0,\pi$,\footnote{We point out 
the vanishing as $(\theta-\pi)^2$ in the smooth limit, and $1/\theta$ divergence in the sharp limit, are expected for CFTs.
The fact that these properties hold in other scale invariant theories is not unreasonable.}
and with the reflection property of pure states, $a(2\pi-\theta)=a(\theta)$. 
% We note that in the smooth limit, $a(\theta)$ vanishes as $\sigma\,(\theta-\pi)^2$ with $\sigma=c/(12\pi^2)$. This result can be put in correspondence
% with the CFT answer, Eq.~\req{conj1}, $\sigma^{\rm CFT}=\pi^2\ctt/24$, where $\ctt$ dictates the energy density 2-point in the groundstate.
% In the Lifshitz theories under consideration, the equal-time correlation function of the energy-density is determined by a 
% correlation function of the parent $d\!=\!2$ CFT 
The smooth limit expansion coefficients, Eq.~\req{expansion}, can be simply obtained by rewriting Eq.~\req{lif} in the form,  
$\tfrac{c}{12}\{-1+1/[1-(\theta-\pi)^2/\pi^2]\}$:
\begin{align} \label{sig-lif}
  \sigma^{(p)}= \frac{c}{12}\, \frac{1}{\pi^{2p+2}}\,.
\end{align}
We recognize a geometric series: the $\sigma^{(p)}$ obey a pure exponential scaling $A/\pi^{2p}$ for all $p$. We also conclude that the series   
has maximal radius of convergence, $\pi$. 
Using our general expression for the asymptotic behavior of $\sigma^{(p)}$ at large $p$, Eq.~\req{sigp-asym}, together with Eq.~\req{sig-lif},    
we predict that the sharp limit coefficient is $\kappa=\pi c/24$. 
A direct expansion of $a(\theta)$ about $\theta=0$ gives $\pi c/(24\theta)$, in agreement with our prediction.
This class of Lifshitz theories thus obeys the smooth-sharp relation that we put forth for CFTs in \S\ref{sec:ssc}.  

Further, we note that Eq.~\req{lif} exceeds the lower bound found for CFTs, Eq.~\req{amin}, when $\amin$ is normalized to have the same
leading smooth-limit coefficient $\sigma$. This follows from the fact that the expansion coefficients of $\amin$ (given in Appendix \ref{ap:exact}) 
are less than those of Eq.~\req{lif}, which can easily be proven by bounding the zeta function. 
Given that the derivation of this lower bound depended on Lorentz invariance, it is not clear how general the result is. For instance,
will more general Lifshitz theories violate the CFT bound?

It would be of interest to relate the expansion coefficients given in Eq.~\req{sig-lif} to correlation functions of local operators, as in the case
of CFTs, for which  the leading coefficient $\sigma$ is given by the stress tensor 2-point function. We leave this task for future investigation. 
Finally, let us mention that corner entanglement has been recently studied for other classes of non-conformal theories using holog
raphy 
% \footnote{Our conjecture \req{conj1} has also been recently used\cite{Dong:2015zba} in the context of holography to 
% provide supporting evidence that total derivative terms in the bulk action do not contribute to the generalized gravitational entropy.}  
\cite{Pang:2015lka,Alishahiha:2015goa,Mozaffar:2015xue}, where significant differences have been found with respect to the CFT results\cite{Bueno1,Bueno2}.   

\section{Higher dimensions}
\label{highd}
\begin{figure}
\center
   \includegraphics[scale=.68]{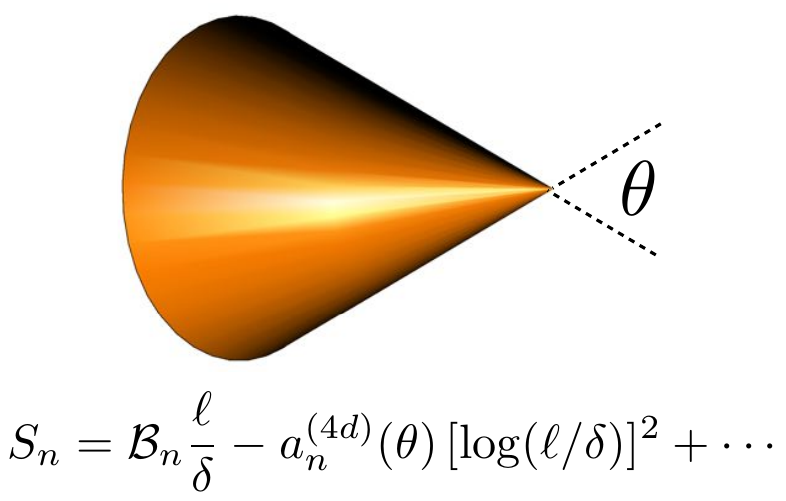}  
   \caption{{\bf Cone entanglement.} A cone singularity in $d=3+1$ contributes logarithmic term to the $n$th
\ren entropy, with a universal prefactor, $-a_n^{(4d)}(\theta)$.
} 
 \label{fig:cone}  
\centering  
\end{figure}

For a conical singularity in $d\!=\! 3+1$, the EE of a CFT receives a subleading logarithmic correction 
as shown in \rfig{fig:cone}. The corresponding cone function $\afour(\theta)$ is independent
of the UV regulator as for corners in $d=3$.  
Interestingly, it is \emph{entirely} fixed by $\ctt$\cite{Klebanov:2012yf}:
\begin{align}\label{a4d1}
 \afour(\theta) = \frac{\pi^4 \ctt}{160}\, \frac{\cos^2(\theta/2)}{\sin(\theta/2)}\,,
\end{align} 
where $\theta$ is the cone's opening angle, see \rfig{fig:cone}.  
%where the smooth limit corresponds again to $\theta=\pi$. 
This expression applies to \emph{all} four-dimensional CFTs.
In addition, the very same $\theta$-dependence holds for the \ren entropies, namely\cite{Bueno4}
\begin{align} \label{a4d1}
 \afour_n(\theta) = \frac{1}{4}f_b(n)\, \frac{\cos^2(\theta/2)}{\sin(\theta/2)}\,,
\end{align} 
where $f_b(n)$ depends on the R\'enyi index\cite{Fursaev:2012mp,Lee:2014xwa}, with $f_b(1)=\pi^4\ctt/40$. Strong evidence\cite{Bueno4,Lee:2014xwa,Lewkowycz:2014jia} suggests that $f_b(n)$ is determined by the twist operator scaling dimension $h_n$ 
through $f_b(n)=(3\pi/2)\, h_n/(n-1)$. (In $d=4$, the twist operator is a surface operator.)
The fact that $h_n$ fully determines the entire $\theta$-dependence is in contrast with $d=3$, 
where the scaling dimension of the twist operator only determines the smooth limit coefficient. 

Many similarities, however, do exist between $d\!=\!4$ and $d\!=\! 3$, and we now discuss them in light of the findings presented above. 
For one, $\afour(\theta)$ admits a smooth limit expansion about $\pi$ in even powers of $(\theta-\pi)$, just as in $d\!=\!3$,
Eq.~\req{expansion}. In Appendix \ref{ap:exact}, we show that this expansion has a radius of convergence of $\pi$. 
We conjecture that is the case for CFTs in all dimensions $d\!>\! 1+1$.     
Further, we show there that the expansion coefficients $\sigma^{(p)}$ are strictly positive, again in agreement with our general conjecture.
Recall that in $d=3$ we have been able to prove this for the first five coefficients, $\sigma^{(p<5)}_n$.  
%Further, in Appendix \ref{ap:exact} we show that the radius of convergence of the expansion is $\pi$, and that 
In \rfig{fig:smoothsharp}, we plot the first 13 coefficients (except the first one, $\sigma$) of the entanglement entropy.
They vanish exponentially with $p$, and rapidly approach the asymptotic form \req{sigp-asym}, \ie $\sigma_n^{(p\gg 1)}\to 2\kappa_n/\pi^{2p+3}$,
where again $\kappa_n$ dictates the small angle divergence of the cone function: $a_n^{(4d)}(\theta\to 0)=\kappa_n/\theta$,
with $\kappa_n=f_b(n)/2$.   

For theories in higher dimensions $d>4$, less is known about the general form of the analogous hypercone function $a^{(d)}_n(\theta)$ away 
from the smooth limit \cite{Myers:2012vs,Safdi:2012sn,Bueno4,Miao2015}, in particular for odd-dimensional theories. For the special case of 
even-dimensional holographic theories, one can perform the calculation, and the properties of the expansion about the smooth
limit are analoguous to what we found here for $d=3$ and 4\cite{pablo-prep}.   

\section{Discussion} \label{sec:disc}   
We have studied the entanglement properties of quantum critical theories.
Focusing on conformal field theories, we have established lower bounds on the corner function $a(\theta)$ in $d\!=\!2+1$, 
Eqs.~(\ref{amin},\ref{inE}), and on its expansion coefficients 
in smooth limit, $\theta\!\to\!\pi$. Our bound for $a(\pi/2)$ corresponding to 90-degree corners, Eq.~\req{bound90}, is nearly saturated by all known
results, including recent numerical estimates on lattice models that describe Wilson-Fisher quantum critical points.   
The bounds rely on fundamental properties of entanglement, such as the strong subadditivity of EE.   
An important extension would be to establish stronger bounds on the corner function for general \ren index than what we obtained, Eq.~\req{an_min}.
It is further natural to ask whether an \emph{upper} bound exists for $a(\theta)$, and its expansion coefficients. 
The holographic correspondence could be helpful in answering this question.

With regards to the physical properties  
of the smooth-limit expansion coefficients, we have shown that the quartic coefficient $\sigma'$ (\rfig{fig:corner}) contains information beyond the 
2- and 3-point functions of the stress tensor,
and conjectured that 4-point data is required. This potentially  makes $\sq$ much richer compared to the leading coefficient $\sigma$ because
4-point functions are highly non-trivial in CFTs. 
It will be interesting to determine the precise physical information encoded in this and higher order coefficients. With such data in hand, 
the bounds we have derived for the corner entanglement function will yield bounds on the physical properties of CFTs. 
Another important direction for future investigation would be to examine the properties of corner entanglement in non-conformal quantum critical
theories, with Lifshitz scaling for instance. Our discussion in \S\ref{sec:lif} suggests that a connection exists between the latter and conformal field
theories, at least for $z=2$ theories with conformal wave functions\cite{ardonne}. 

% \textbf{\emph{Note added}} --- Remarkably, soon after the appearance of this paper, our original conjecture \req{conj1} was proven for completely general CFTs in Ref.~\onlinecite{Faulkner:2015csl}. The results in that paper also imply that Mezei's formula\cite{Mezei:2014zla} for the entanglement entropy of slightly deformed hyperspherical entangling surfaces in $d$-dimensional CFTs is valid for general theories and, consequently, that the generalized version of our conjecture for entangling surfaces containing hyperconical singularities\cite{Bueno4} is true for general CFTs. The generalized version of our conjecture for arbitrary values of the R\'enyi index (both for three-dimensional theories \req{conj2} and in higher dimensions\cite{Bueno4}) remains unproved in general though.
\begin{acknowledgments}   
We are thankful to H.~Casini and R.\ C. Myers for many stimulating discussions. 
We also acknowledge useful exchanges with A.~Cueto, O.~Lasso, R.~Melko, R.-X.~Miao, P.~F.~Ram\'irez and E.~M.~Stoudenmire. 
WWK is funded by a fellowship from NSERC. The work of PB was supported
by a postdoctoral fellowship from the Fund for Scientific Research - Flanders (FWO) and partially by the COST Action MP1210 The String Theory Universe.
The work of WWK was in part performed at the Aspen Center for Physics, which is supported by 
National Science Foundation grant PHY-1066293.
\end{acknowledgments}

\onecolumngrid  
\appendix
\section{Establishing the lower bounds}
\label{ap:bounds}

\subsection{Proving the minimality of $\amin(\theta)$} 
\label{ap:proof}
We shall prove that $a(\theta) \geq \amin(\theta)$, where
the minimal function $\amin$ is given in closed form in Eq.~\req{amin}. 
To do so, we rederive a special instance of a theorem proved by
S.A.~Chaplygin (see page 164 of Ref.~\onlinecite{cloud2014}). 

Let us consider the differential inequality:
\begin{align} \label{pf1}
  \ddot a(\theta) \geq F(\theta,\dot a(\theta))\,,
\end{align} 
where dots denote $\theta$ derivatives.
We are interested in the particular case where $F=-\dot a/\sin\theta$, but the proof below holds more generally. 
We shall work on the interval $\pi\leq \theta < 2\pi$, where $\dot a \geq 0$.
The minimal function $\amin$ satisfies the differential equation obtained by replacing 
the inequality in \req{pf1} with an equality, \ie
\begin{align} \label{pf2}
  \ddot{\frak a}_{\rm min}(\theta) = F(\theta,\dot{\frak a}_{\rm min}(\theta))\,.
\end{align}
Both functions satisfy the same boundary conditions at $\theta=\pi$:
\begin{align}
  a(\pi) &=\dot a(\pi) = 0\,, \nn
  \amin(\pi) &=\dot{\frak a}_{\rm min}(\pi) = 0 \,.
\end{align}
Thus the leading term in the Taylor expansion around $\pi$ is $\sigma\,(\theta-\pi)^2$,
with $\sigma$ taken to be the same for both functions.
Subtracting \req{pf2} from \req{pf1}, we get
\begin{align} \label{pf3}
  \ddot a - \ddot {\frak a}_{\rm min} - (\dot a -\dot{\frak a}_{\rm min})\, Q(\theta) \geq 0
\end{align}
where we have defined the quotient
\begin{align} \label{pf4}
  Q(\theta) = \frac{F(\theta,\dot a)- F(\theta,\dot{\frak a}_{\rm min})}{\dot a-\dot{\frak a}_{\rm min}}\,,
\end{align} 
which is well-defined even at points where $\dot a=\dot{\frak a}_{\rm min}$. The key step is to introduce 
the new function
\begin{align} \label{pf5}
  U(\theta) = \exp\left[-\int_\pi^\theta \!d\vartheta \, Q(\vartheta) \right]\,, 
\end{align}
which is non-negative. We can thus rewrite \req{pf3} as: 
\begin{align} \label{pf6}
  \partial_\theta\left\{ (\dot a-\dot{\frak a}_{\rm min}) U(\theta) \right\} \geq 0\,.
\end{align} 
Integrating both sides of this inequality between $\pi$ and $\theta>\pi$, and using $\dot a(\pi)=\dot{\frak a}_{\rm min}(\pi)$, we obtain
\begin{align} \label{pf7} 
  \dot a(\theta) \geq \dot{\frak a}_{\rm min}(\theta)\,,
\end{align}
for $\pi\leq \theta<2\pi$. (By virtue of the reflection property of $a(\theta)$ about $\pi$,
this leads to $|\dot a(\theta)|\geq |\dot{\frak a}_{\rm min}(\theta)|$ for $0\leq \theta \leq \pi$.)
Finally, we note that
\begin{align} \label{pf8}
  a(\theta) = \int_\pi^\theta d\theta\, \dot a(\theta)\,,
\end{align}
and the same holds for $\amin$. We can thus integrate \req{pf7} to obtain the desired result:
\begin{align}
  a(\theta) \geq \amin(\theta) \,. 
\end{align}
This completes our proof for the lower bound on the corner function. 

\subsection{Bounds for general \ren index}
\label{ap:ren-bound}
We now give the derivation of the lower bound for $a_n(\theta)$, Eq.~\req{an_min}, following a very similar 
approach as in the previous subsection. We begin by re-writing the third inequality obtained from reflection 
positivity, Eq.~\req{RP2}: 
\begin{align} \label{RP2a}
  \partial^4_\theta a_n\times [\partial_\theta^2 a_n] \geq (\partial_\theta^3 a_n)^2 \,, 
\end{align}
where the term in square brackets is non-negative by virtue of the second reflection positivity constraint, Eq.~\req{an_convex}.
This implies that we can write the inequality in the form $\partial_\theta^4 a_n\geq G$, where $G$ depends only on the second 
and third derivatives of $a_n$. We can now apply a higher order version of Chaplygin's theorem invoked above\cite{cloud2014}. 
Clearly,
\begin{align}
  \aminn(\theta) = \sigma_n\, (\theta-\pi)^2\,,
\end{align}
is a solution of the differential equation obtained by replacing the inequality in Eq.~\req{RP2a} by an 
equality. For convenience, we shall work on the interval $[\pi,2\pi)$. The theorem guarantees that 
\begin{align}
  a_n(\theta) \geq   \aminn(\theta)\,, 
\end{align}
assuming they have matching initial value conditions at $\pi$:
\begin{align}
  a_n(\pi) &= \aminn(\pi) =0\,, \nn
  \partial_\theta a_n(\pi) &= \partial_\theta \aminn(\pi) = 0\,, \nn
  \partial_\theta^2 a_n(\pi) &= \partial_\theta^2 \aminn(\pi) = \sigma_n \,.
\end{align}
This proves the desired lower bound, Eq.~\req{an_min}.

We now show that all the reflection positivity inequalities Eq.~\req{RP}, labelled by an integer $M\geq 1$, lead to lower bounds, and
never to an upper bound. Let us denote the $M$th inequality by $I_M\geq 0$. We then have
\begin{align}
  \partial_\theta^{2M} a_n(\theta) \times I_{M-1} + P_M \geq 0\,,
\end{align}
where $P_M$ is a linear combination of products of derivatives of $a_n$.
Since $I_{M-1}\geq 0$, we have:
\begin{align} \label{rp-ode}
  \partial_\theta^{2M} a_n(\theta) \geq G_M \,,
\end{align}
where $G_M$ only depends lower order derivatives of $a_n$.
Applying Chaplygin's theorem as above, one will obtain a lower bound for $a_n$ by 
solving the non-linear differential equation Eq.~\req{rp-ode}. In this case one needs
to supply $(M-1)$ initial conditions at $\theta=\pi$. We leave the investigation of the resulting inequalities
for future work.
 
\section{Exact smooth limit expansions} \label{ap:exact}
We determine the smooth limit expansion in closed-form for the corner function $a_n(\theta)$
in a variety of systems. We begin with the minimal corner function $\amin(\theta)$, Eq.~\req{amin}, then
analyze the Extensive Mutual Information model for CFTs, and finally the cone function describing \emph{all} CFTs 
in $d=3+1$. In all cases, we will find that the radius of convergence  
of the Taylor expansion about $\theta=\pi$ is $\pi$. The series must breakdown at $\theta=0$
since the the corner function has a $1/\theta$ divergence as $\theta\to 0$. 
In relation to the latter, we find that in all cases (except for $\amin$, which is not 
the corner function of an actual CFT, but rather a bound) the expansion coefficients asymptote to
\begin{align} \label{sig_asympt}
  \sigma_n^{(p)} \xrightarrow{p\gg 1} \frac{2}{\pi^3}\, \frac{\kappa_n}{\pi^{2p}}\,, 
\end{align}
thus decaying exponentially fast at large $p$, as explained in \S\ref{sec:ssc}. The coefficient $\kappa_n$ dictates the small
angle divergence,
\begin{align}
  a_n(\theta\to 0) = \kappa_n / \theta\,. 
\end{align}

\subsection{Minimal function}
The minimal function, Eq.~\req{amin},
\begin{align} \label{amin2}
  \amin(\theta) = \frac{\pi^2\ctt}{3} \, \log\left[ 1/\sin(\theta/2) \right]\,,
\end{align}
has the smooth limit expansion
\begin{align}
  \amin(\theta) =\sum_{p=1}^\infty \sigma_{\rm min}^{(p-1)}\,(\theta-\pi)^{2p}\,,
\end{align}
where
\begin{align} \labell{amin-coeff}
  \sigma^{(p-1)}_{\rm min} = \frac{\pi^2\ctt}{3}\; \frac{(1-2^{-2p})\,\zeta(2p)}{p\,\pi^{2p}}\,,
\end{align}
$\zeta(z)$ is Riemann's zeta function. 
Since $\zeta(2p)$ is strictly positive for all integers $p\geq 1$, all the coefficients are 
positive. %; in fact, $\sigma^{(p)}_{\rm min}>8/(p+1)$
We note that $\zeta(2p)/\pi^{2p}$ is a rational number related to the Bernoulli number $B_{2p}$: $\zeta(2p) = (-1)^{p+1}(2\pi)^{2p} B_{2p}/[2(2p)!]$.  
We list the leading coefficients in Table \ref{tbl_min}. We can use the ratio test to check the convergence of the series. One finds
\begin{align}
\lim_{p\rightarrow \infty}\left| \frac{\sigma_{\rm min}^{p+1}\, (\theta-\pi)^{2(p+2)}}{\sigma_{\rm min}^{p}\,(\theta-\pi)^{2(p+1)}} \right| = \frac{(\theta-\pi)^2}{\pi^2}\, ,
\end{align}
so the series is convergent when $(\pi-\theta)^2/\pi^2<1$, \ie for all $\theta \in (0,\pi]$.
At large $p$, the coefficients decay to zero as   
\begin{align}
  \sigma_{\rm min}^{(p)} \xrightarrow{p\gg 1} \frac{\ctt}{3}\; \frac{1}{p\,\pi^{2p}}\,,
\end{align}
where the extra factor of $p$ in the denominator implies that $\amin(\theta)$ decays faster 
compared to what is expected of a CFT, Eq.~\req{amin2}.  

\begin{table*}
  \centering
  \begin{tabular}{c||c|c|c|c|c|c} 
    $p$ & 0 & 1 & 2 & 3 & 4 & 5 \\ \hline \hline \rule{0pt}{1.5em}
    $\sigma_{\rm min}^{(p)}\, 24/(\pi^2 \ctt) $ & 1 & $1/24$ & $1/360$  & $17/80640$ & $31/1814400$ & $691/479001600$ 
  \end{tabular} 
  \caption{Smooth limit coefficients $\sigma_{\rm min}^{(p)}$ for the minimal function $\amin(\theta)$.
}
\labell{tbl_min}    
\end{table*}    

\subsection{Extensive Mutual Information model}
We here analyze the corner function of the Extensive Mutual Information (EMI) model\cite{Casini:2008wt,Swingle:2010jz,Casini:2005rm}.
As the name suggests, it is characterized by the property that the mutual 
information, $I(A,B)=S(A)+S(B)-S(A\cup B)$, satisfies the extensivity property\cite{Casini:2008wt,Casini:2005rm}:
$I(A,B\cup C)=I(A,B)+I(A,C)$. 
The corner function in the EMI model reads:\cite{Casini:2008wt,Swingle:2010jz,Casini:2005rm}
\begin{align}\label{aemi1}
  a^{\rm emi}_n(\theta) = \frac{3h_n}{\pi(n-1)}\,\, [1+(\pi-\theta)\cot\theta]\,,
\end{align} 
where $h_n$ is the scaling dimension of the twist operator. This normalization was first obtained in Ref.~\onlinecite{Bueno1}. 
We impose on $h_n$ all the conditions required for a CFT, in particular $h_n/(n-1)\geq 0$. This ensures that $a^{\rm emi}_n(\theta)$
is positive. Taylor expanding about $\theta=\pi$, 
\begin{align}
  a^{\rm emi}_n(\theta) =\sum_{p=1}^\infty \sigma_{n}^{(p-1)}\,(\theta-\pi)^{2p}\,,
\end{align}
we obtain the smooth limit coefficients in closed form:
\begin{align}
 \sigma_{n}^{(p-1)}= \frac{3h_n}{\pi(n-1)}\,\frac{2\zeta(2p)}{\pi^{2p}}  \,.
\end{align}
By virtue of the positivity of the zeta function (see previous subsection), all these coefficients are strictly positive.  
Further, at $n=1$ the first coefficient, $\sigma$, is equal to the one given by the minimal function $\amin(\theta)$, 
whereas all the higher order coefficients are strictly greater:
\begin{align}
 \sigma^{(p)} > \sigma_{\rm min}^{(p)}\,, \qquad \text{for }\; p \geq 1\,,
\end{align}
where we have again omitted the subscript $n=1$. 
We list the first few coefficients in Table~\ref{tbl_emi}. We note that this series is convergent 
for all $\theta \in (0,\pi]$. % \begin{align}
%   \sigma_n^{1}\frac{\pi(n-1)}{h_n} &= 1 \nn
%   \sigma_n^{2}\frac{\pi(n-1)}{h_n} &= \frac{1}{15}
% \end{align}

The $p\gg 1$ asymptotic behavior of the coefficients reads:
\begin{align}
  \sigma^{(p)} \xrightarrow{p\gg 1} \frac{\ctt}{4}\; \frac{1}{\pi^{2p}}=\frac{2\kappa}{\pi^{2p+3}}\,,   
\end{align}
vanishing exponentially fast and in agreement with Eq.~\req{sigp-asym}.  

\begin{table*}
  \centering
  \begin{tabular}{c||c|c|c|c|c|c} 
    $p$ & 0 & 1 & 2 & 3 & 4 & 5 \\ \hline \hline \rule{0pt}{1.5em}
    $\sigma_{n}^{(p)}\cdot \pi(n-1)/h_n$ & 1 & $1/15$ & $2/315$ & $1/1575$ & $2/31185$ & $1382/212837625$
  \end{tabular} 
  \caption{Smooth limit coefficients $\sigma_{n}^{(p)}$ for the corner function of the Extenstive Mutual Information model. 
}
\labell{tbl_emi}    
\end{table*}

\subsection{Cones in $4d$ CFTs}
As we argued in  \S\ref{highd}, in three spatial dimensions ($d\!=\!4$), a conical singularity in the entangling surface gives rise to
the following universal function\cite{Klebanov:2012yf,Bueno4}:
\begin{align}
  \afour_n(\theta) = \frac{3\pi h_n}{8(n-1)} \frac{\cos^2(\theta/2)}{\sin(\theta/2)}\,, 
\end{align} 
where the smooth limit again corresponds to $\theta=\pi$. Just as above, $h_n$ is the scaling
dimension of the twist operator (which has support on two dimensional surfaces in $4d$).
In that limit $\afour$ has the Taylor expansion
\begin{align}
  \afour_n(\theta)=\sum_{p=1}^{\infty} \sfour^{(p-1)}\, (\theta-\pi)^{2p} \, .
\end{align}
The coefficients are given by
\begin{align}
  \sfour^{(p-1)} = \frac{3\pi h_n}{8(n-1)}\, \frac{(-1)^p(E_{2p}-1)}{4^{p}(2p)!} \, , 
\end{align}
where $p\!\geq\! 1$, and $E_{2p}$ is the Euler number. We note that $E_{2p}= (-1)^p A_{2p}$, 
where $A_{2p}$ is the so-called zig-zag Euler number, which is a positive integer
with some combinatorical significance.
We thus conclude that all the expansion coefficients are strictly positive. 
Further, just as in all the cases studied above, the radius of convergence of the series is $\pi$, 
namely the expansion converges on $(0,2\pi)$. To establish this one needs to use the fact that
$|E_{2(p+1)}/E_{2p}|\to 16p^2/\pi^2$ as $p\to\infty$, which can be readily derived from the asymptotic
behavior of $E_{2p}$ at large $p$: $E_{2p}\to (-1)^p 8\sqrt{p/\pi} \left( \tfrac{4p}{\pi e} \right)^{2p}$. 

The large $p$ behavior of the coefficients is
\begin{align}
  \sfour^{(p)} \xrightarrow{p\gg 1} \frac{6h_n}{\pi^2(n-1)}\; \frac{1}{\pi^{2p}}=\frac{2\kappa_n}{\pi^{2p+3}}\,,
\end{align}
where we again find the exponential decay given by $1/\pi^{2p}$. 

%\subsection{Cones in $6d$ CFTs}
%In five spatial dimensions, a hyperconical singularity in the entangling surface generically produces a universal contribution controlled by a function of the opening angle $a_n^{6d}(\theta)$ which is analogous to the four dimensional one. As opposed 

\section{Stress tensor 2- and 3-point functions}  \label{ap:ttt} 
The stress or energy-momentum tensor $T_{\mu\nu}(x)$ has scaling dimension $d$, equal to
the spacetime dimension. In a CFT, its groundstate 2- and 3-point functions are strongly
constrained by conservation of energy and momentum, and by the conformal symmetry. 
In this appendix, we review their basic properties.
The 2-point function reads\cite{Osborn} 
\begin{align} \label{TT2}
  \langle T_{\mu\nu}(x) T_{\eta\kappa}(0)\rangle = \frac{\ctt}{x^{2d}} \, \mathcal I_{\mu\nu,\eta\kappa}(x)\,, 
\end{align}
where 
\begin{align}
  \mathcal I_{\mu\nu,\eta\kappa}(x) = \frac{1}{2}\left( I_{\mu \eta}(x)I_{\nu\kappa}(x)+I_{\mu\kappa}(x)I_{\nu\eta}(x)\right)
  -\frac{1}{d}\delta_{\mu\nu}\delta_{\eta\kappa}\,,
\end{align}
where we have defined the dimensionless tensor $I_{\mu\nu}(x)=\delta_{\mu\nu}-2x_\mu x_\nu/x^2$, and we are working in Euclidean time. 
From Eq.~\req{TT2}, we see that the 2-point function is entirely determined by a single real number in all dimensions, $\ctt\geq 0$.
In Table~\ref{tbl-TTT}, we list the value of $\ctt$ in free boson and free Dirac fermion CFTs, and CFTs holographically dual 
to pure Einstein gravity.  
\begin{table*} 
  \centering
  \begin{tabular}{c||c|c|c}
   $\;$  & scalar & fermion & AdS/CFT\\ \hline\hline  
   $\ctt$ & $3/(16\pi^2)$ & $3/(16\pi^2)$  & $3L^2/(\pi^3G)$ \\ \hline
   $t_4$ & $4$ & $-4$ & 0 
  \end{tabular}
  \caption{Stress tensor 2- and 3-point function parameters $\{\ctt,t_4\}$ in $d=3$ for a complex scalar, Dirac fermion, 
and CFTs holographically dual to pure Einstein gravity. 
}
\label{tbl-TTT}   
\end{table*}  

The stress tensor 3-point function $\langle T_{\mu\nu}(x_1) T_{\eta\kappa}(x_2)T_{\sigma\rho}(0)\rangle$ is more complex,
and its explicit form can be found in Ref.~\onlinecite{Osborn}. 
The main point is that it is entirely characterized by only two parameters in $d=3$\cite{Osborn}. 
(In $d\geq 4$, a third parameter is needed\cite{Osborn}.)
A Ward identity\cite{Osborn} relating 
the 2- and 3-point functions of the stress tensor implies that one of these parameters can be chosen to be $\ctt$, encountered above.
We choose the second parameter to be $t_4$\cite{buchel}. Table~\ref{tbl-TTT} gives the value of $t_4$ for the free and holographic CFTs.
% For instance, in Einstein holography, these take the values:
% \begin{align}
%   \ctt = \frac{24 L^2}{\pi^2 \ell_P^2}\,, \qquad \quad
% %  t_2 &= 0\nn %24(\lagb + 3\mu)\nn
%   t_4 = 0\,,
% \end{align}
% where the Planck length squared is given by $\ell_P^2=8\pi G$. 
For general CFTs, $\ctt$ and $t_{4}$ are related to the couplings defined by Osborn and Petkou\cite{Osborn}, 
$\bl a,\bl b,\bl c$, appearing in $\langle TTT\rangle$ as follows: 
\begin{align}
  \ctt &= 4S_d \frac{(d-2)(d+3)\bl a -2\bl b -(d+1)\bl c}{d(d+2)}\,, \nn
  % t_2 &= \frac{2 (d+1) \left[ (d-1) \left(d^2+8 d+4\right) \bl a +3 d^2\bl b -(2 d+1) d\, \bl c \right]}{d \left[
  %  (d-2) (d+3)\bl a -2 \bl b- (d+1)\bl c \right]} \nn
  t_4 &=-\frac{(d+1) (d+2) \left[ 3 (d-1) (2 d+1)\bl a +2 d^2\bl b - d(d+1) \bl c \right]}{d \left[
   (d-2) (d+3)\bl a -2 \bl b- (d+1)\bl c \right]}\,,   \label{ctt4}
\end{align}
where $S_d=2\pi^{d/2}/\Gamma(d/2)$. The first equation follows from a Ward identity of the stress tensor.
We emphasize that $\ctt t_4$ is a \emph{linear} combinations of the couplings $\bl a,\bl b,\bl c$.
This is important because these latter couplings are additive for free bosons and fermions\cite{Osborn}, meaning that  
$\ctt t_4$ is additive for these theories. 
% \begin{align}
%  \ctt= 4S_d \frac{(d-2)(d+3)\bl a -2\bl b -(d+1)\bl c}{d(d+2)} \,.
% \end{align}
% \comment{The above equation appears twice}
For $d=3$, Eq.~\req{ctt4} reduces to:
\begin{align}
  \ctt &= \frac{32\pi}{15}(3\bl a-\bl b -2\bl c)\,, \nn
%  t_2 \ctt &= \frac{128\pi}{45} (74 \bl a + 27 \bl b - 21 \bl c)  \nn
 t_4\ctt  &= -\frac{128\pi}{3} (7 \bl a + 3 \bl b - 2 \bl c)\,.
\end{align}
Using these equations we can for instance solve for the Osborn-Petkou parameters for CFTs holographically 
dual to pure Einstein gravity:  
\begin{align}
  \frac{\bl a}{\ctt} = -\frac{27}{256 \pi}\,, \qquad  %\frac{27 [-1 + 16 (\lagb+3\mu)]}{256 \pi} \nn
  \frac{\bl b}{\ctt} = -\frac{3}{256 \pi}\,, \qquad  %-\frac{3  [1 + 144 (\lagb+3\mu)]}{256 \pi} \nn
  \frac{\bl c}{\ctt} = -\frac{99}{256 \pi}\,. %\frac{9 [-11 + 96 (\lagb+3\mu)]}{256 \pi}
 \end{align}

\section{Proof of the inequality $\sigma_n^{(4)}\geq 0$}
\labell{ineq56}

In \S\ref{sec:smooth} we showed that on general grounds,  
\begin{align}\labell{sig4}
\sigma_n^{(4)} &\geq \frac{45(\sigma_n'')^3-168\sigma_n'\sigma_n''\sigma_n'''+392\sigma_n(\sigma_n''')^2}{126[5\sigma_n\sigma_n''-2(\sigma_n')^2]}\, .
\end{align}
The denominator is always positive by virtue of the second equation in \reef{smooth-Ren-ineq}, so we only need to prove the positivity of the numerator.
Using the second and third inequalities in \reef{smooth-Ren-ineq}, we define two quantities, $k_1$ and $k_2$,  
\begin{align}
k_1&=\sigma_n''-\frac{2}{5}\frac{(\sigma_n')^2}{\sigma_n}\geq 0\, ,\\
k_2&=\sigma_n'''-\frac{15}{28}\frac{(\sigma_n'')^2}{\sigma_n'}\geq 0\, ,
\end{align}
which are non-negative. Substituting $\sigma_n''$ and $\sigma_n'''$ in terms of these constants in \reef{sig4}, we find
\begin{multline} \labell{sig41}
45(\sigma_n'')^3-168\sigma_n'\sigma_n''\sigma_n'''+392\sigma_n(\sigma_n''')^2=392 k_2^2\sigma_n+\\
\frac{3k_1}{10 \sigma_n^2(\sigma_n')^2} \left[5 k_1 \sigma_n+2 (\sigma_n')^2\right] \left[75 k_1^2\sigma_n^2+60 k_1
  \sigma_n (\sigma_n')^2 + 280 k_2 \sigma_n^2 \sigma_n'+12 (\sigma_n')^4\right] \,, 
\end{multline} 
which is a sum of non-negative terms. This implies $\sigma_n^{(4)}\geq 0$.  

\bibliography{quartic-refs}{}    
\end{document}